\RequirePackage{fix-cm}
\documentclass[twocolumn]{svjour3}          
\smartqed  
\usepackage{graphicx}
\usepackage{amssymb}
\usepackage{bm}
\usepackage{hyperref}
\usepackage{mathrsfs} 
\usepackage[intlimits]{amsmath}
\usepackage{todonotes}
\usepackage{changes}
\definechangesauthor[name=YS,color=brown]{YS}
\definechangesauthor[name=KW,color=purple]{KW}
\definechangesauthor[name=YL,color=orange]{YL}
\usepackage{float}
\usepackage{subfigure}
\usepackage{mathptmx}      
\journalname{SN Applied Sciences}
\graphicspath{{Arxiv_Figure/}}
\DeclareMathOperator{\trace}{tr}
\DeclareMathOperator{\dev}{dev}

\begin{document}

\title{A Manifold Learning Approach to Accelerate Phase Field Fracture Simulations in the Representative Volume Element \thanks{This work is supported by the Guangdong Province Key Area R\&D Program, grant No.~2019B010940001, by the Shanghai Natural Science Foundation, grant No.~19ZR1424200, and by the National Natural Science Foundation of China, grant No.~11972227.}
}


\author{Yangyuanchen Liu  \and
        Kexin Weng  \and
        Yongxing Shen
}

\authorrunning{Liu, Weng, and Shen} 

\institute{Y. Liu 
           \and
           K. Weng 
           \and
           Y. Shen \at
           University of Michigan -- Shanghai Jiao Tong University Joint Institute \\
           Shanghai Jiao Tong University \\
           800 Dongchuan Road, Shanghai, 200240, China\\
              \email{yongxing.shen@sjtu.edu.cn}         \\     
              \\
              K. Weng \at
College of Engineering\\
University of Michigan\\
Ann Arbor, MI 48109-2102, USA
}

\date{Received: date / Accepted: date}

\maketitle

\begin{abstract}
The multiscale simulation of heterogeneous materials is a popular and important subject in solid mechanics and materials science due to the wide application of composite materials. However, the classical FE$^2$ (finite element$^2$) scheme can be costly, especially when the microproblem is nonlinear.  In this paper, we consider the case when the microproblem is the phase field formulation for fracture.  We adopt the locally linear embedding (LLE) manifold learning approach, a method for non-linear dimension reduction, to extract the manifold that contains a collection of phase-field-represented initial microcrack patterns in the representative volume element (RVE). Then the output data corresponding to any other microcrack pattern, e.g., the evolved phase field at a fixed load, can be accurately reconstructed using the learned manifold with minimum computation. The method has two features: a minimum number of parameters for the scheme, and an input-specific error bar. The latter feature enables an adaptive strategy for any new input on whether to use the proposed, less expensive reconstruction, or to use an accurate but costly high-fidelity computation instead.
\end{abstract}


\section{Introduction}
\label{intro}
Heterogeneous materials such as composites have been widely applied in various industries such as aircraft and automobile manufacturing.
The multiscale simulation of heterogeneous materials is therefore a crucial task in computational mechanics. 

\begin{figure}[htpb]
	\centering
	\includegraphics[width=\columnwidth]{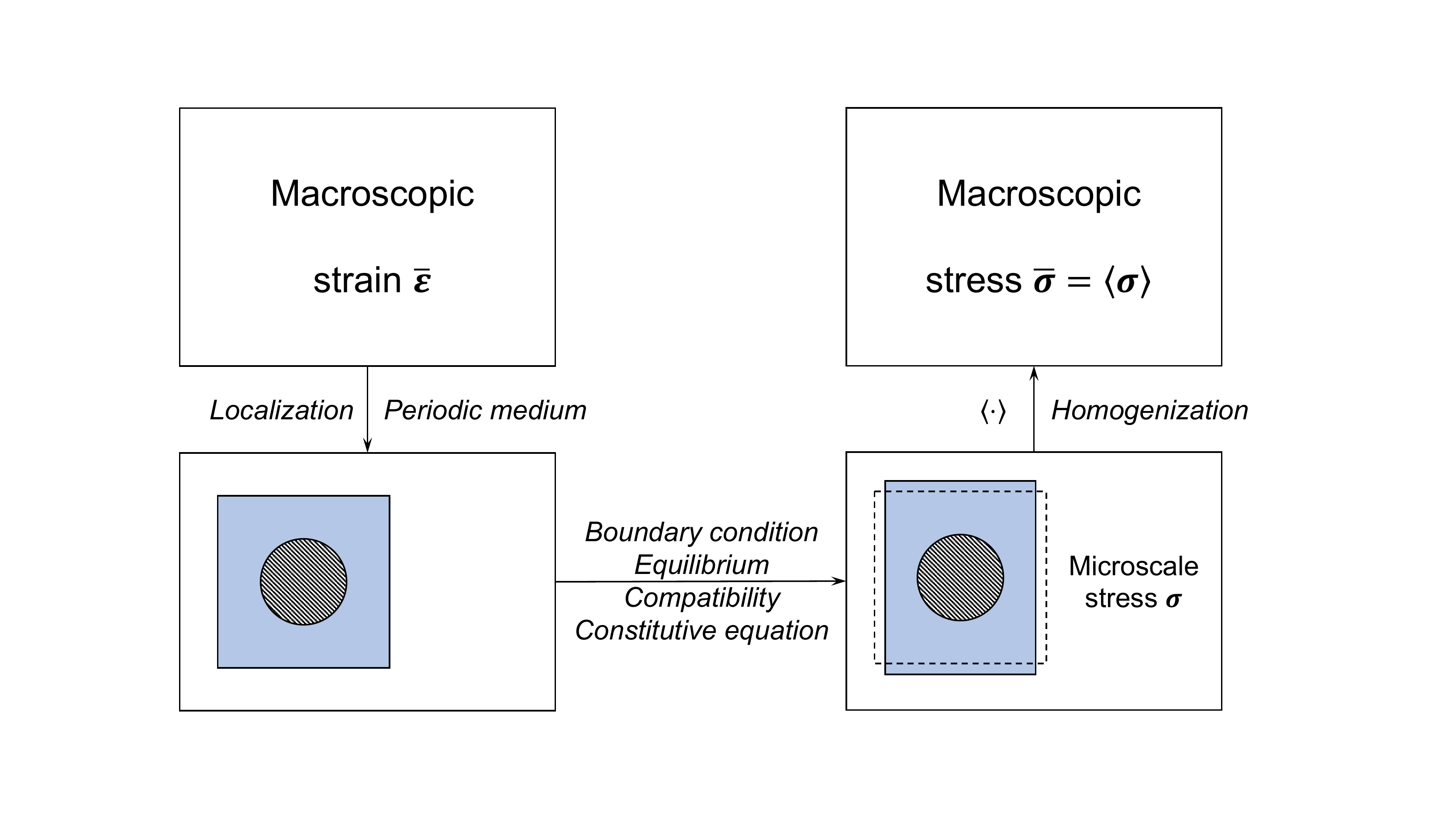}
	\caption{Flowchart illustrating the FE$^2$ scheme.}
	\label{fig:fe2scheme}
\end{figure}

Such simulation is usually facilitated by the classical FE$^2$ scheme \cite{feyel1999}, as illustrated in Fig.~\ref{fig:fe2scheme}. 
In a typical FE$^2$ scheme, the finite element method is applied at the microscale and the macroscale concurrently, and hence the name. More precisely, at the macroscale, the entire composite part is discretized into continuum finite elements, each of which has several Gauss quadrature points for numerical integration. For each Gauss quadrature point, the effective constitutive behavior for the macroscale is obtained through a homogenization process via a finite element analysis at the microscale. The computational domain at the microscale is called a representatixve volume element (RVE).  
Take the mechanical simulation for a fiber-reinforced composite as an example, a typical RVE consists of a fiber and the surrounding matrix \cite{Lee2020112694}, possibly with defects such as cracks. Normally the desired effective responses include the stress tensor and the elasticity tensor, and the simplest way of homogenization is by volume averaging.




Among available numerical methods for the analysis at the RVE with crack propagation, the phase field approach to fracture \cite{bourdin2000numerical}, also known as the regularized variational theory for fracture, shows clear advantages. This approach is built on Griffith’s theory for brittle fracture \cite{griffith1921}. The key idea is to use a scalar field, called phase field, to represent the crack path, instead of incorporating the explicit geometry of the crack path in the computational domain. The advantages include obviating the need for explicitly tracking the crack path geometry, and the ability to predict crack nucleation and bifurcation without extra criterion. The method has since been applied to fracture modeling in Euler-Bernoulli beams \cite{LAI2020}, thin shells \cite{AMIRI2014}, composite materials \cite{ZHANG2019105008,ZHANG2020111551}, cement-based materials \cite{NGUYEN20191}, layered structures \cite{NGUYEN2019585}, and CO$_2$ fracturing \cite{MOLLAALI2019}.

However, solving the equations arising from the phase field method for fracture can be costly. Since the strain energy functional to minimize in this approach is not convex, the required number of iterations for convergence is not known \emph{a priori}. The RVE analysis is, of course, no exception. Many efforts have been devoted to accelerating the phase field fracture solution procedure.
Heister \emph{et al.}~\cite{HEISTER2015466} and Li \emph{et al.}~\cite{li2019} constructed mesh adaptivity approaches for the problem.
Ziaei-Rad and Shen \cite{vahid2016} developed a massively parallel algorithm for the phase field approach with time adaptivity.
Gerasimov and De Lorenzis \cite{GERASIMOV2016276} proposed a line search procedure for the monolithic scheme to overcome the iterative convergence issues of non-convex minimization.
Wick \cite{wick2017a,wick2017b} developed modified Newton-Raphson schemes for fully monolithic quais-static brittle phase field fracture propagation.
Farrell and Maurini \cite{farrell2017} reformulated the staggered algorithm of the phase field analysis as a nonlinear Gauss-Seidel iteration and employed over-relaxation to accelerate convergence.
Wu \emph{et al.}~\cite{WU2020112704} developed a quasi-Newton monolithic methodwith the Brodyen-Fletcher-Goldfarb-Shanno (BFGS) algorithm. 
Kopani\v{c}\'{a}kov\'{a} and Krause \cite{KOPANICAKOVA2020} proposed a trust region method with application to monolithic phase-field fracture models. 

We aim to accelerate the multiscale simulation from another perspective.
In fact, in many cases, the RVEs are similar within the same multiscale analysis. This similarity can be exploited to accelerate computation, for example, via manifold learning. 	

In the machine learning context, manifold learning is employed to extract the manifold that represents high-dimensional data points and to perform data reconstruction with a minimum amount of computation. Manifold learning has been widely applied to multiscale analysis \cite{satyaki2016,wirtz2015,wang2013,yvonnet2007}, see also the review by Matou\v{s} \emph{et al.}~\cite{MATOUS2017}.
An instance of manifold learning techniques is locally linear embedding (LLE).
Proposed by Roweis and Saul \cite{roweis2000}, LLE is an unsupervised learning algorithm that computes low-dimensional, topology-preserving embeddings of high-dimensional data points. As an instance of kernel principal component analysis (kernel PCA), LLE has many attractive properties. For example, 
the local geometry of high-dimensional data is preserved in the low-dimensional manifold. LLE is particularly suitable for problems with a large amount of similar high-dimensional data.

However, LLE assumes that the data all reside on a single continuous manifold \cite{chen2011}, which poses certain restrictions on the application. For example, in image-based simulations \cite{lopez2018}, each RVE is represented as a vector containing, e.g., pixel values. 
In this case, if the dimension of this vector varies between RVEs, the nonuniform data structure will make LLE training and interpolation impossible. 
This is because the neighborhood finding and interpolation operations of the LLE algorithm requires that the linear combination of data points to be well defined.

Despite such restrictions, the advantages of LLE make it ideal for random RVE computation and computational homogenization \cite{lopez2018,Dolbow2019,Ibanez2018} for multiscale analysis of heterogeneous materials. 


Inspired by \cite{lopez2018} for heat conduction problems,
for the problem of multiscale fracture simulation at hand, 			
we aim to learn a manifold that contains a collection of similar cracked RVEs, 
 and to efficiently compute any desired output dependent on such microstructure using LLE reconstructions. 
Concretely speaking, the input is chosen as the phase field pattern at the beginning of a certain time step (termed ``initial phase field'' for short), and the output can be the phase field at the end of the time step -- so as to make a closed loop for the analysis of the next step -- or any other derived quantity from such phase field solution such as the homogenized stress. In the discrete picture, we construct a finite element mesh to describe the RVE, interpolate the phase field for the crack pattern using the finite element basis functions, and vectorize the description of the initial crack pattern of each RVE using the nodal values of the phase field. The desired output is the phase field solution corresponding to a certain boundary condition. 

Compared with recent contributions on applying machine learning techniques, neural networks in particular, for constitutive modeling \cite{Furukawa1998,Ghaboussi1998,Hashash2004,JUNG2006,SUN2010,JI2011} and similar computations for RVEs \cite{Jiang2014,asahi2017,ren2018,WEI2018908,LI2019735}, the adopted method possesses the following features. 

First, the number of hyperparameters is minimal: only the size of the neighborhood and the number of reduced dimensions need to be input by the user. The selection of such hyperparameters is determined by a systematic cross-validation approach. 

Second, there is no limit on the dimension of the desired output, as long as it is a continuous functional of the microstructure, while a typical neural network would have one set of thresholds and weights per scalar output. 

Finally, for any new input, the uncertainty (``error bar'') for the reconstructed output can be obtained, as a strong correlation is observed between the reconstruction error and a parameter solely dependent on the input information. In this case, the parameter is the distance from the new input to the learned data manifold. This last feature enables a criterion to be developed to assess the reconstruction error \emph{a priori}; in other words, a criterion to decide whether to use the reconstruction which is less expensive, or resort to the high-fidelity computation which is more accurate. This also serves as an indicator of whether the collection of inputs should be augmented with the new input in question, in a greedy sampling fashion, should some kind of adaptivity is to be implemented.  

However, it is still worth noting that, just like many other machine learning techniques, the LLE approach requires enough data points to guarantee the accuracy of predictions. Hence the training set should be dense and large enough. Moreover, as inherited from the general LLE technique,
the proposed approach requires the data structure to be homogeneous, making the distance function and linear combination between data points well-defined. Finally, the output should continuously depend on the input data, which is also a necessary condition for a well-posed problem anyway.

The content of this paper is structured as follows. In Section \ref{fe2}, the FE$^2$ scheme and phase field method are introduced. In Section \ref{manifold}, the manifold learning and LLE techniques are explained in detail. In Section \ref{numerical}, numerical implementations and results are illustrated with error assessments. Finally, in Section \ref{conclusions}, a summary of the proposed computational strategy is presented.
\section{The FE$^2$ Scheme Applied to Composite Fracture}
\label{fe2}

In this section, we introduce the FE$^2$ scheme in the multiscale fracture simulation of a fiber-reinforced composite. The FE$^2$ is a two-scale modeling scheme which applies FE discretizations at both macro and micro scales, the former taking input from the latter through the analysis of the RVE.

\begin{figure}[htpb]
	\centering
	\includegraphics[width=\columnwidth]{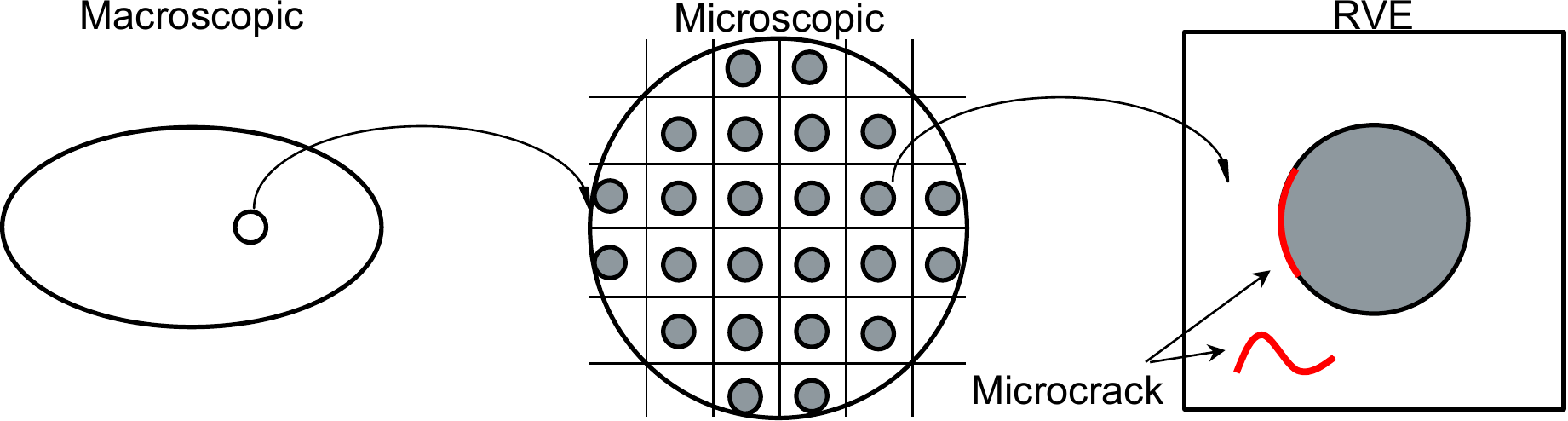}
	\caption{Modeling a macroscopic composite as a collection of RVEs.}
	\label{fig:rve_multisacle}
\end{figure}

In our case, as shown in Fig.~\ref{fig:rve_multisacle}, the RVE is composed of a strong fiber in the center with a weaker matrix. We aim to perform the fracture simulation of the cracked RVE at the microscale. Once the local behavior is determined, the overall macroscopic response of the RVE can be obtained using any well-established homogenization theory and be used for the macroscopic simulation.

\begin{figure}[htpb]
    \centering
    \includegraphics[width=1\columnwidth]{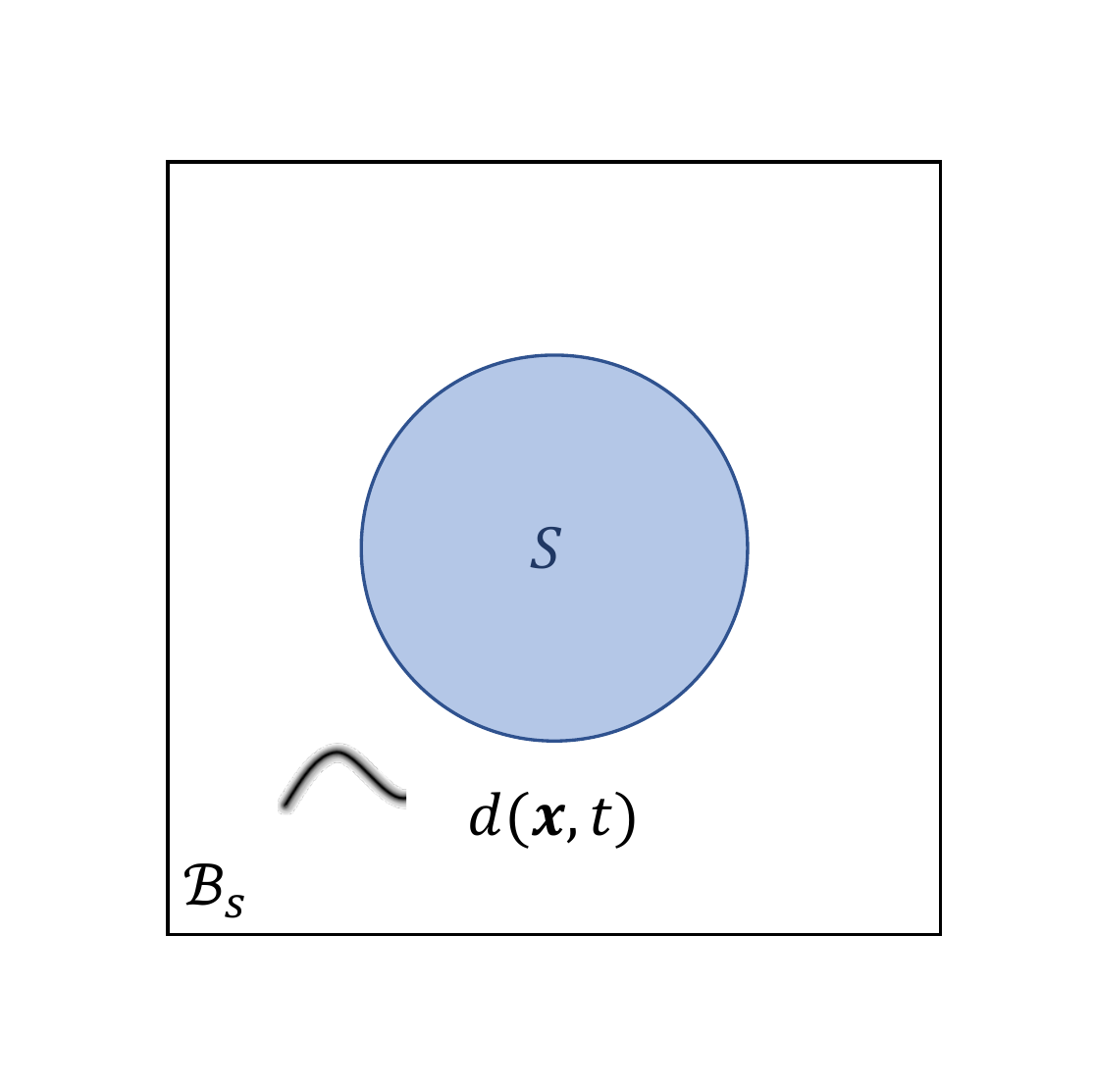}
    \caption{The simplified RVE to be analyzed in this work. In this RVE there is a strong fiber inside a weaker matrix. The only allowed form of failure is matrix cracking.}
    \label{fig:rve_micro}
\end{figure}

For simplicity, we only consider the microcrack evolution in the matrix and ignore all other defects, such as cracks on the interface (debonding) and in the fiber, see Fig.~\ref{fig:rve_micro}.


\paragraph
{Phase Field Approach for RVE Cracking.}
Among many crack simulation methods, we adopt the phase field method to simulate the microcrack evolution in RVE. The phase field modeling of brittle fracture has shown its advantages on simulating complex fracture process, such as obviation of remeshing, see \cite{bourdin2000numerical,amor2009regularized,miehe2010_1273}. The phase field approach of fracture is based on the variational energy formulation proposed by \cite{francfort1998revisiting}, which can be considered as a generalization of Griffith’s theory \cite{griffith1921}. 

\begin{figure}[htpb]
    \centering
    \subfigure[]{\label{subfig:pd_illus_1}\includegraphics[width=0.3\columnwidth]{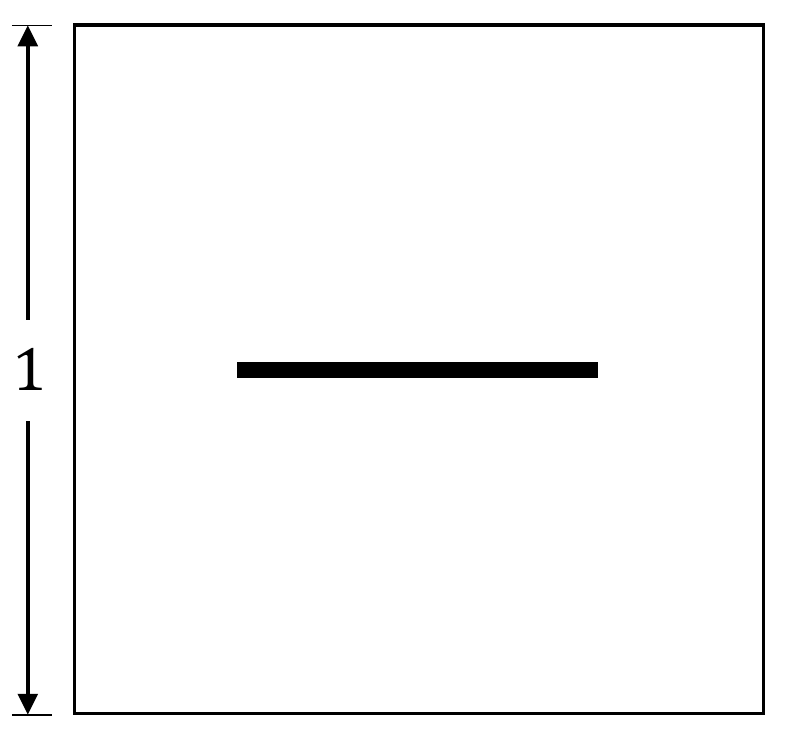}}
    \subfigure[]{\label{subfig:pd_illus_2}\includegraphics[width=0.35\columnwidth]{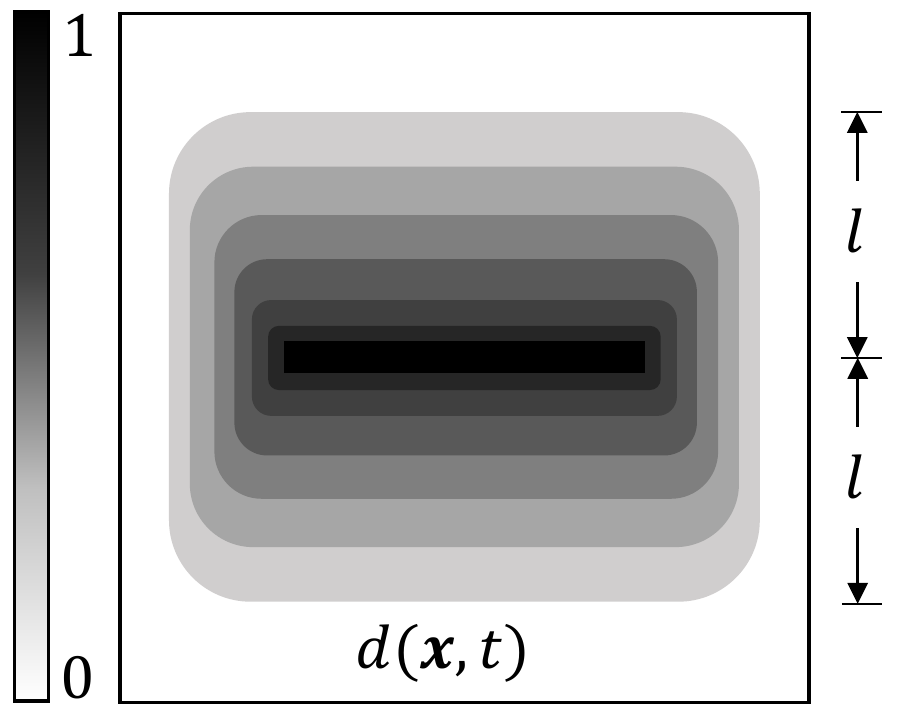}}
    \subfigure[]{\label{subfig:pd_illus_3}\includegraphics[width=0.28\columnwidth]{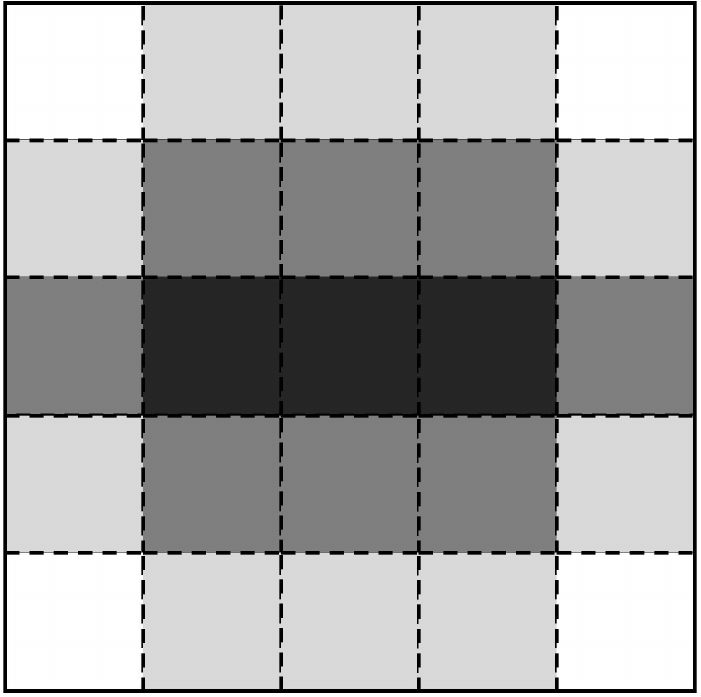}}
    \caption{Representations of a unit cracked microstructures: (a) discrete crack model; (b) phase field corresponding to (a); (c) pixel representation of the phase field model with a structured quadrilateral mesh of with $h/l=0.5$.}
    \label{fig:pd_illus}
\end{figure}

As shown in Fig.~\ref{subfig:pd_illus_2}, the phase field method uses a diffuse field $d$ to represent the cracked microstructures where $d=0$ represents the intact material and $d=1$ the crack. Then equipped with a finite element mesh, cracked microstructures can be represented as a vector containing the nodal values of the phase field, and the distance of the cracked RVEs can be measured as the Euclidian norm of the difference of such vectors.


Compared with a geometric description of cracks [Fig.~\ref{subfig:pd_illus_1}] which may require a heterogeneous data structure (such as the coordinates of a possibly varying number of discrete points on the evolving crack), the phase field method is advantageous in terms of data structure for the manifold learning approach, as each cracked microstructure can be uniformly represented as a vector consisting of the nodes' phase field values. 
This feature is favorable in the manifold learning process introduced in Section \ref{manifold}, as we can adopt a data structure for the inputs (and outputs) as vectors of the same length. 

\begin{figure}[htpb]
    \centering
    \subfigure[Phase field (chosen)]{\label{subfig:pd_func}\includegraphics[width=\columnwidth]{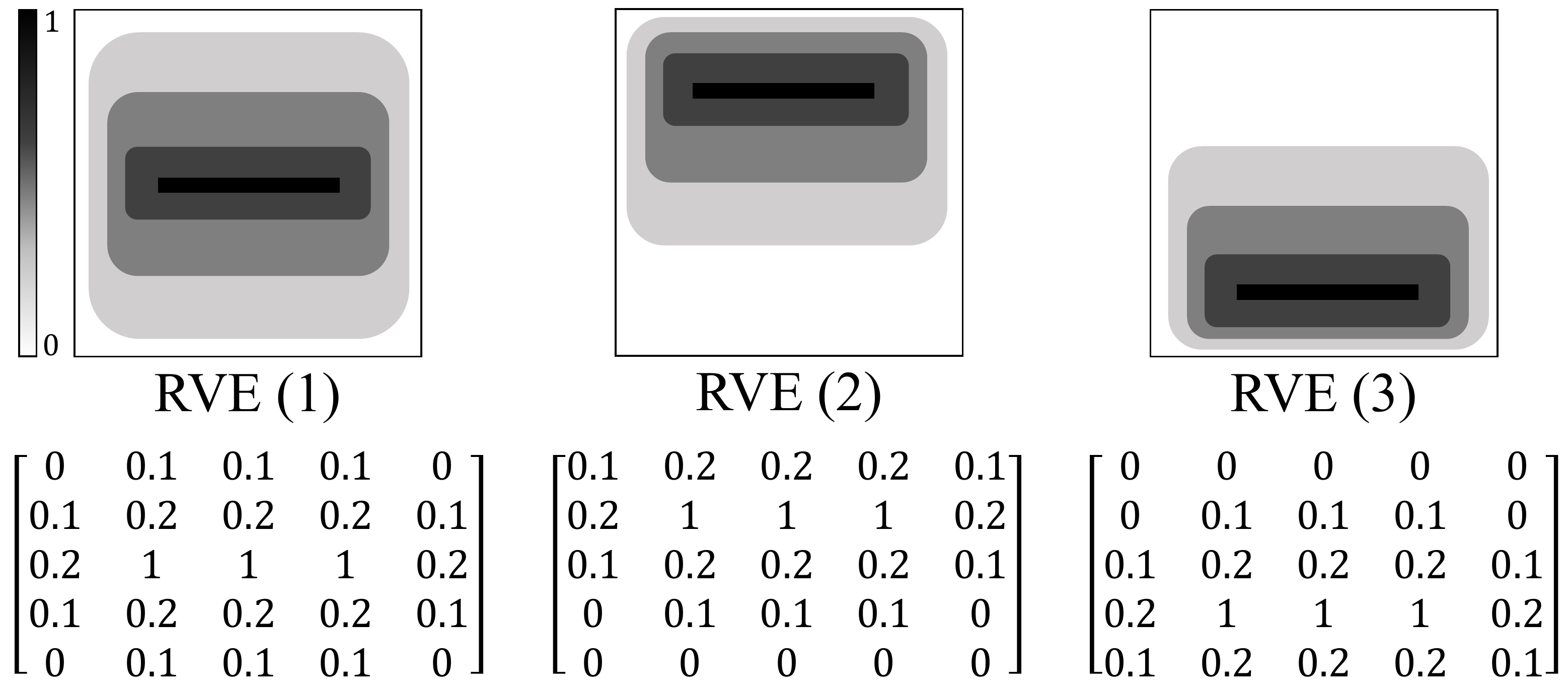}}
    \subfigure[Characteristic function (not recommended)]{\label{subfig:char_func}\includegraphics[width=\columnwidth]{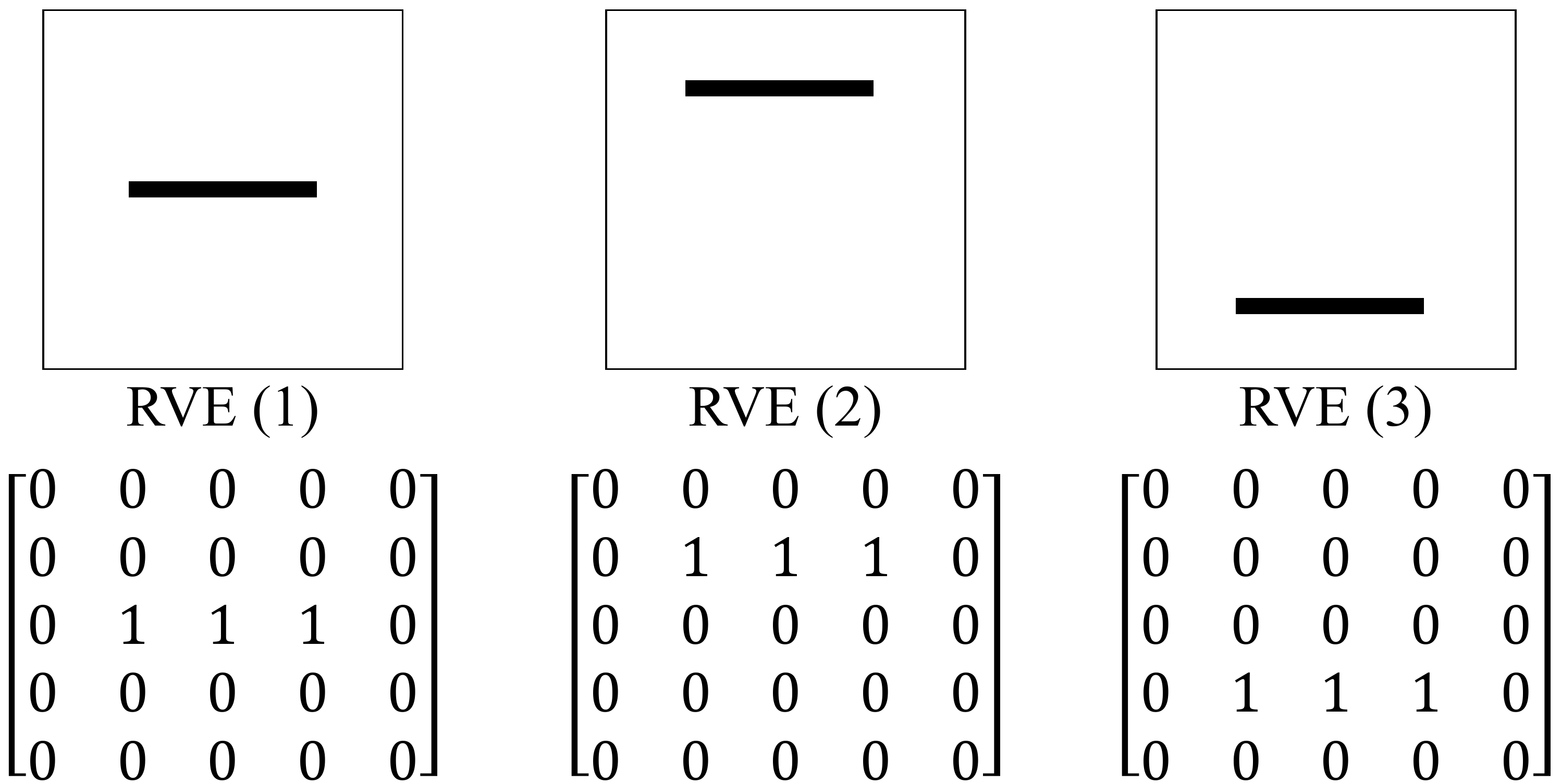}}
    \subfigure[Distance function (not recommended)]{\label{subfig:dist_func}\includegraphics[width=\columnwidth]{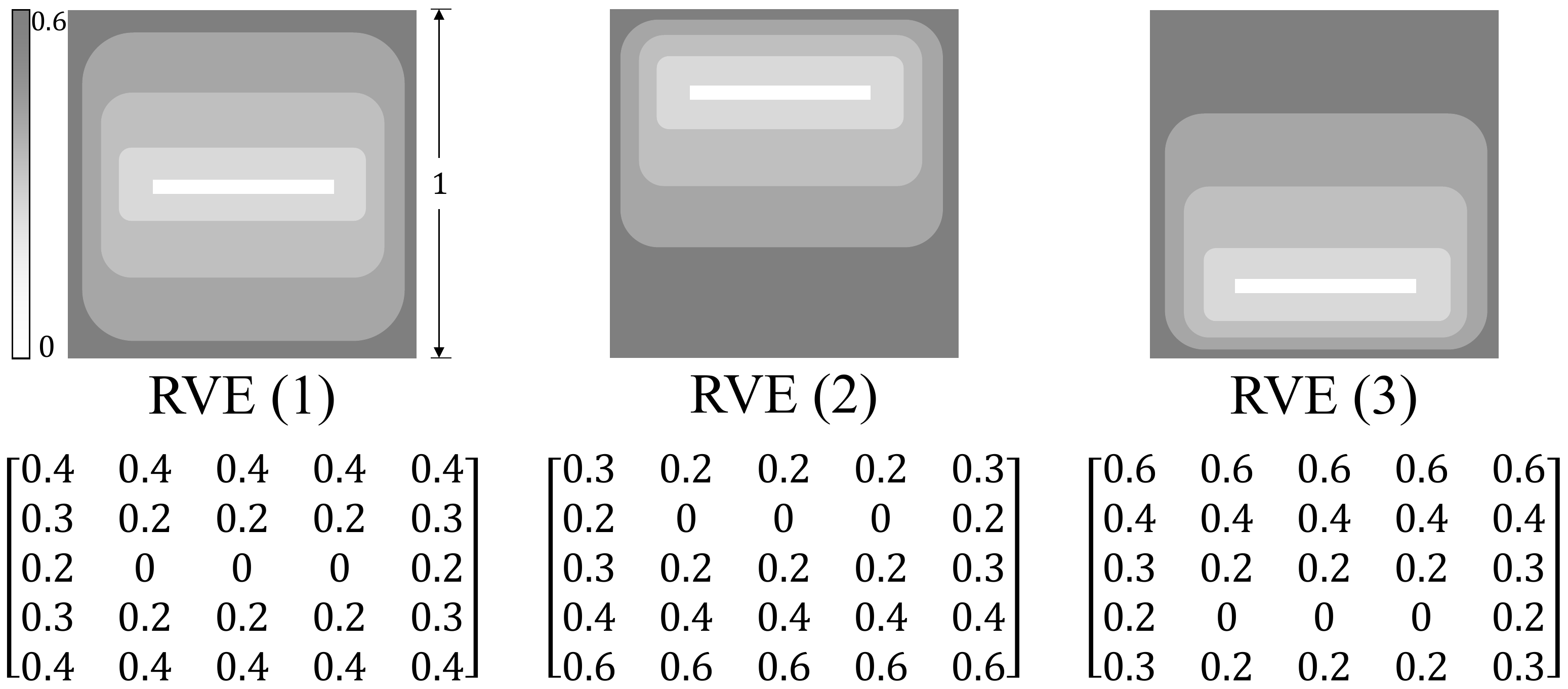}}
    \caption{Numerical representations of cracked microstructures with 5$\times$5 nodes: (a) phase field (chosen); (b) characteristic function (not recommended); (c) distance function (not recommended). We employ (a) since it is able to vectorize cracked microstructures, and the distance metric between crack patterns is well defined. (b) would not be able to tell the distance of non-overlapping cracks and (c) would weight too much on the difference of crack pattern pairs in areas \emph{far away} from the cracks.}
    \label{fig:crack_func}
\end{figure}

Fig.~\ref{subfig:pd_illus_3} and Fig.~\ref{subfig:pd_func} 
show the pixel representations of microcracks with the phase field approach.
At the first sight, there are at least two possible alternatives to translate a cracked microstructure into a numerical representation: (1) using the characteristic function of the cracks, i.e., 1 for the crack and 0 otherwise, as shown in Fig.~\ref{subfig:char_func}, (2) using the distance function to the cracks, as shown in Fig.~\ref{subfig:dist_func}. Considering that we will need to quantify the ``distance'' of such microstructures, both alternatives present severe drawbacks:
method (1) would not be able to tell the distance of non-overlapping cracks, while method (2) would weight too much on the difference of crack pattern pairs in areas \emph{far away} from the cracks.

The adopted variant
of the phase field formulation is as follows.
In a plane strain setting, let $\mathcal{B}=(-L,L)^2$ be the area initially occupied by the RVE. Within the RVE, let $S\subset\subset\mathcal{B}$ be the fiber, and $\mathcal{B}_s=\mathcal{B}\setminus\overline{S}$ be the matrix, see Fig.~\ref{fig:rve_micro}.
In the absence of body force and traction boundary condition, 
the phase field formulation for the RVE is \cite{shen2018}
\begin{equation}
    \label{eq:variational}
    \begin{split}
        \Pi_{l}[\bm{u},d]  = 
        \int_{\mathcal{B}_s}\varPsi\left [\bm{\varepsilon},d \right ]   \mathrm{d}\mathcal{B} + \int_{S}\varPsi_1(\bm{\varepsilon})   \mathrm{d}\mathcal{B} \\
        + \frac{g_{c}}{2}\int_ {\mathcal{B}_s}\left ( \frac{d^2}{l} + l\left | \nabla d \right |^2 \right )\mathrm{d}  \mathcal{B},
    \end{split}
\end{equation}
where the arguments $\bm{u}\in H^1(\mathcal{B},\mathbb{R}^2)$ and $d\in H^1(\mathcal{B}_s)$ are the displacement field and the phase field, respectively, and the strain tensor is defined as $\bm{\varepsilon}=(\nabla\bm{u}+\nabla\bm{u} ^T)/2$. Here we set the convention for the phase field $d$ as $d=1$ represents the crack and $d=0$ the intact material. Let $(\lambda,\mu)$ and $(\lambda_1,\mu_1)$ be the Lam\'e constants of the matrix and the fiber, respectively, then the strain energy density for the fiber is given by
\[\varPsi_1(\bm{\varepsilon}) = \frac{\lambda_1}{2}(\trace \bm{\varepsilon})^2 + \mu_1 \bm{\varepsilon}:\bm{\varepsilon},\]
while that for the matrix also depends on $d$, for which we adopt the formulation proposed by Amor \emph{et al.}~\cite{amor2009regularized}. This model splits the strain energy density $\varPsi$ into volumetric and deviatoric parts: 
\[\varPsi(\bm{\varepsilon},d)=g(d)\varPsi_+(\bm{\varepsilon}) + \varPsi_-(\bm{\varepsilon}),\]
where
\begin{subequations}\label{eq:v-d}
	\begin{align}
	&\varPsi_+(\bm{\varepsilon}) = \frac{K}{2} \left< \trace\bm{\varepsilon}\right>^2_+ + \mu\|\dev \bm{\varepsilon}\|^2,\\
	&\varPsi_-(\bm{\varepsilon}) = \frac{K}{2} \left< \trace\bm{\varepsilon}\right>^2_-,\\
	&\bm{\sigma}(\bm{\varepsilon}, d) = g(d)\left(K\left< \trace\bm{\varepsilon}\right>_+\bm{1} + 2\mu \dev\bm{\varepsilon}\right) + K \left< \trace\bm{\varepsilon}\right>_-\bm{1}, 
	\end{align}
\end{subequations}
where $K=\lambda + 2\mu/3$ is the bulk modulus, $\dev \bm{\varepsilon} := \bm{\varepsilon} - (1/3)(\trace \bm{\varepsilon})\bm{1}$, $\left<a\right>_{\pm}:=(a\pm|a|)/2$ and the degradation function $g(d) = (1 - d)^2 +k$, where $k$ is a small positive number.
The positive numbers $g_c$ and $l$ are the energy release rate of crack propagation and the regularization length scale, respectively. 



The strong form of the governing equations, except the displacement boundary condition at $\partial\mathcal{B}$, read
\begin{subequations}\label{eq:rve-bvp}
    \begin{align}
        &\text{div } \bm{\sigma}  = \bm{0}, \quad \text{in } \mathcal{B}_s \cup S, \label{eq:RVEeqm}\\
        &\bm{\sigma} = \frac{\partial \varPsi}{\partial \bm{\varepsilon}}, \quad \text{in } \mathcal{B}_s,\label{eq:constitutive1}\\
        &\bm{\sigma} = \frac{\partial \varPsi_1}{\partial \bm{\varepsilon}}, \quad \text{in } S,\label{eq:constitutive2}\\
        &\frac{\partial \varPsi}{\partial d} g_c \left( \frac{d}{l} - l\Delta d \right) = 0, \quad \text{in } \mathcal{B}_s, \label{eq:strongform}\\
        &\bm{\sigma}\cdot \bm{n} \big|_{\mathcal{B}_s} = \bm{\sigma}\cdot \bm{n} \big|_S \quad \text{on } \partial S \label{eq:stress_continuity}\\
        &\bm{u} \big|_{\mathcal{B}_s} = \bm{u} \big|_{S} \quad \text{on } \partial S\label{eq:displacement_continuity}\\
        &\nabla d \cdot \bm{n} = 0\text{ on } \partial \mathcal{B}_s,
    \end{align}
where $\bm{n}$ denotes the outward unit normal vector of $\partial S$ or $\partial \mathcal{B}$.


The general quasi-static calculation 
 for each load step of the microcrack evolution is shown in Fig.~\ref{fig:rvebc}: the inputs are the crack configuration (represented by a phase field) at time $t$ and the boundary conditions for $\bm{u}$ and $d$ at the next time step $t+\Delta t$, and the output is the updated phase field at $t+\Delta t$.  Here $t$ represents a time-like variable to indicate the process of load increment, and likewise $t+\Delta t$.


\begin{figure}[htpb]
    \centering
    \includegraphics[width=\columnwidth]{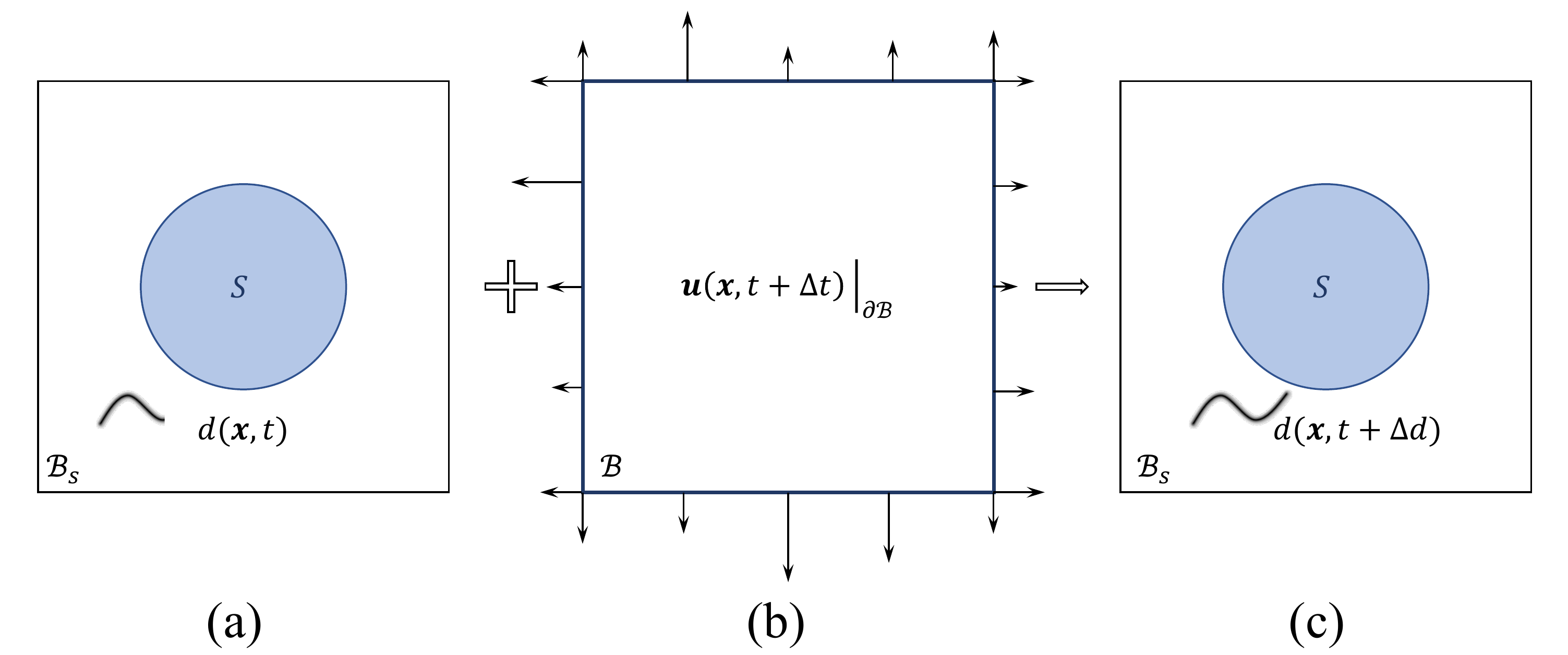}
    \caption{(a) RVE with micro cracks; (b) the boundary conditions of RVE; (c) RVE with evolved micro cracks}
    \label{fig:rvebc}
\end{figure}

 For simplicity, we \emph{fix} the following boundary conditions on $\partial\mathcal{B}$ and focus on the effect of the crack path at $t$ on its updated counterpart at $t+\Delta t$.
Let $\overline{\bm{\varepsilon}}\in\mathbb{R}^{2\times2}$ be the imposed macroscopic strain tensor, then the boundary conditions are set to be
\begin{align}
        &\bm{u} = \overline{\bm{\varepsilon}} \cdot \bm{x}, \quad \text{on } \partial\mathcal{B}.\label{eq:boundary}
\end{align}
\end{subequations}

\section{Manifold Learning Details}
\label{manifold}

The FE$^2$ scheme introduced in Section \ref{fe2} requires an unpredictable number of iterations for convergence due to the non-convexity of the functional $\Pi_l$. 
In order to reduce computational cost, we adopt the so-called manifold learning method. 
The manifold learning scheme uses techniques traditionally designed for machine learning purposes to extract the manifold that represents high-dimensional data points and perform reconstruction with minimum amount of computation \cite{lopez2018,Ibanez2018}. 
The main idea is to generate enough inputs and pre-compute their outputs offline, in this case the phase fields at $t$ and $t+\Delta t$, respectively, then provides the desired output for any input by reconstruction.

In this section, we will elaborate on the manifold learning approach and the LLE technique \cite{roweis2000}, specialized to the problem stated in Section \ref{fe2}. 
In particular, as we fix the load shown in Fig.~\ref{fig:rvebc}(b), the only input to consider is the initial crack path (i.e. the initial phase field) [Fig.~\ref{fig:rvebc}(a)], and the output is the evolved phase field [Fig.~\ref{fig:rvebc}(c)] upon equilibrium. 

\subsection{Locally Linear Embedding}
\begin{figure}[htpb]
    \centering
    \subfigure[]{\includegraphics[width=\columnwidth]{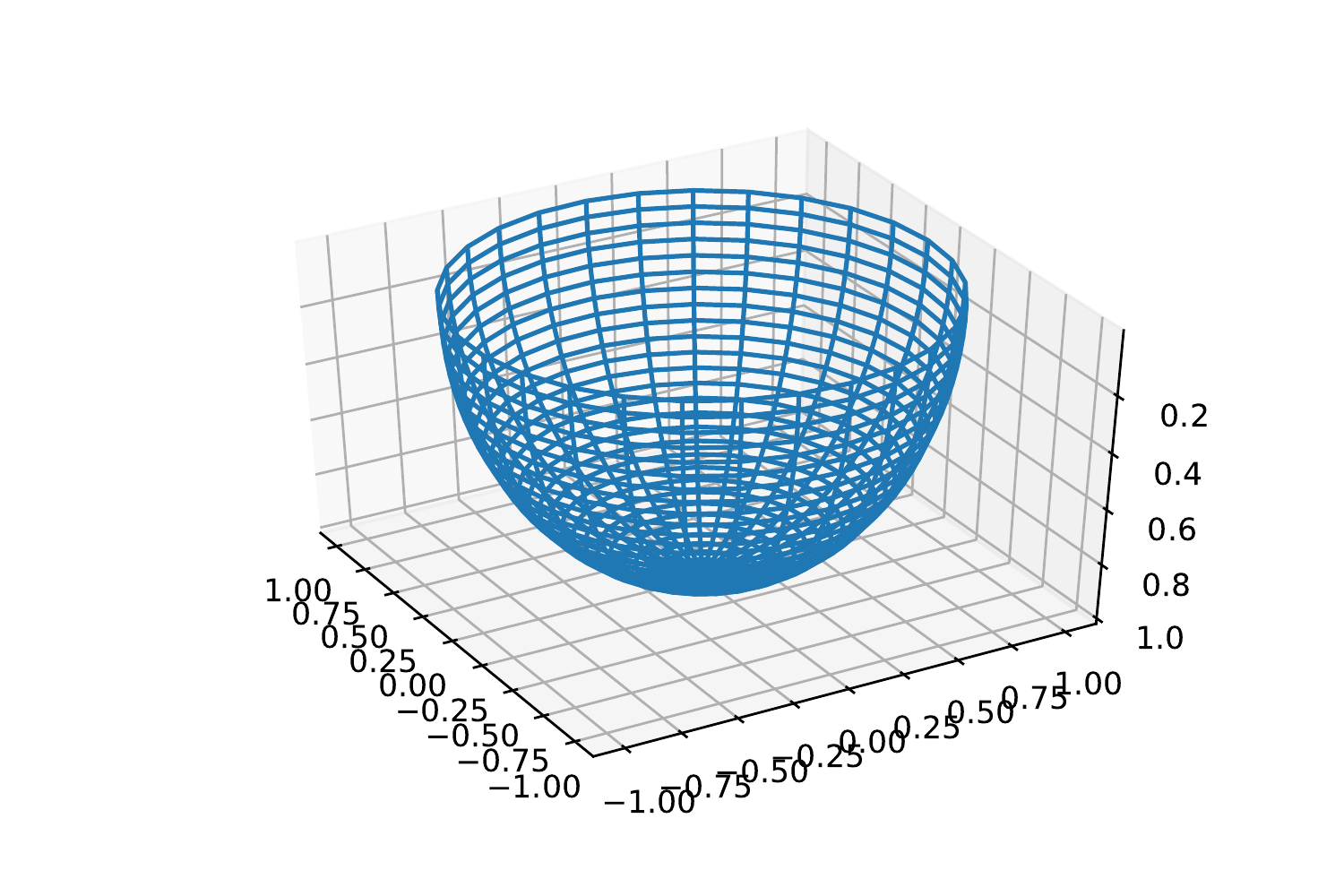}}
    \subfigure[]{\includegraphics[width=0.8\columnwidth]{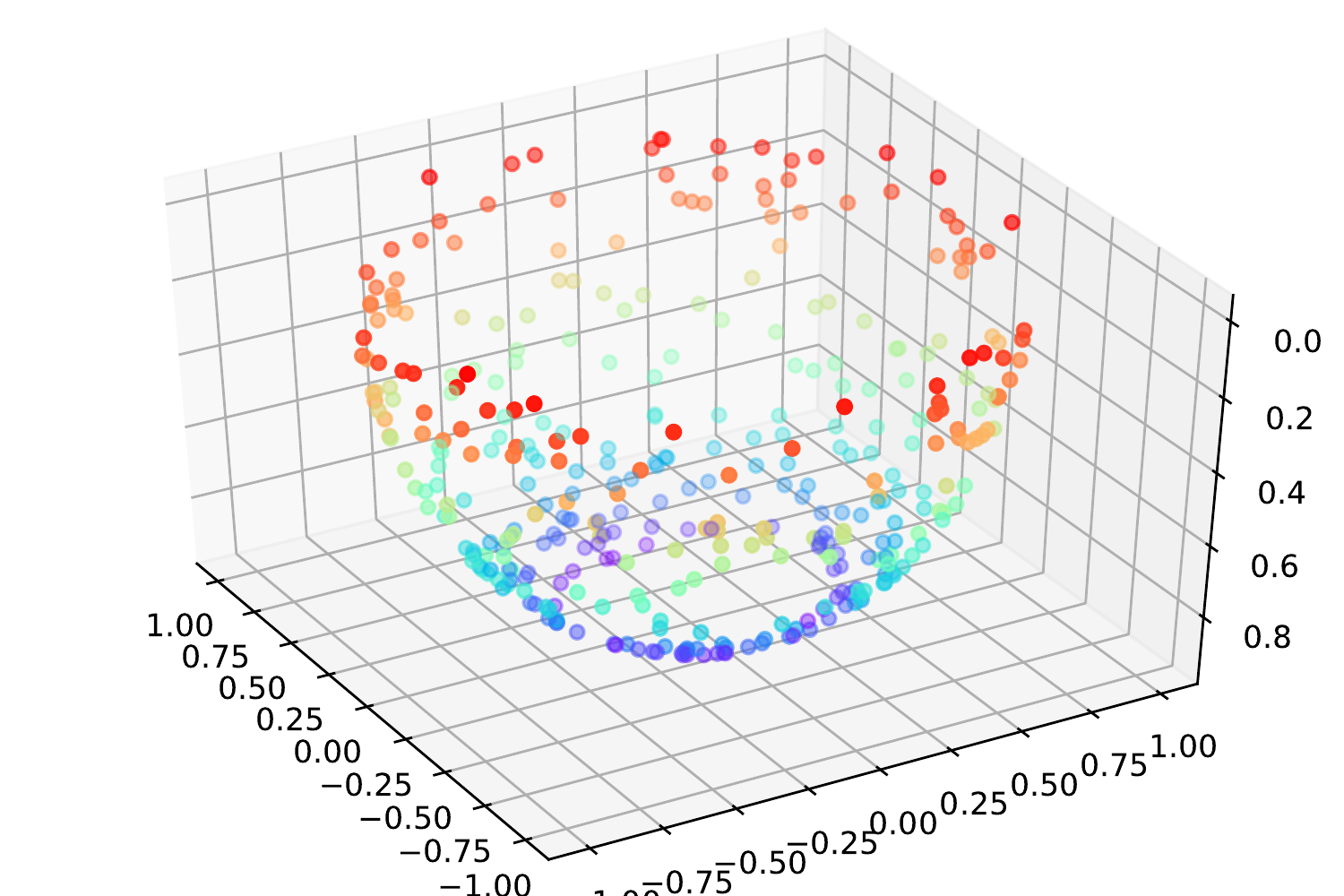}}
    \subfigure[]{\includegraphics[width=0.8\columnwidth]{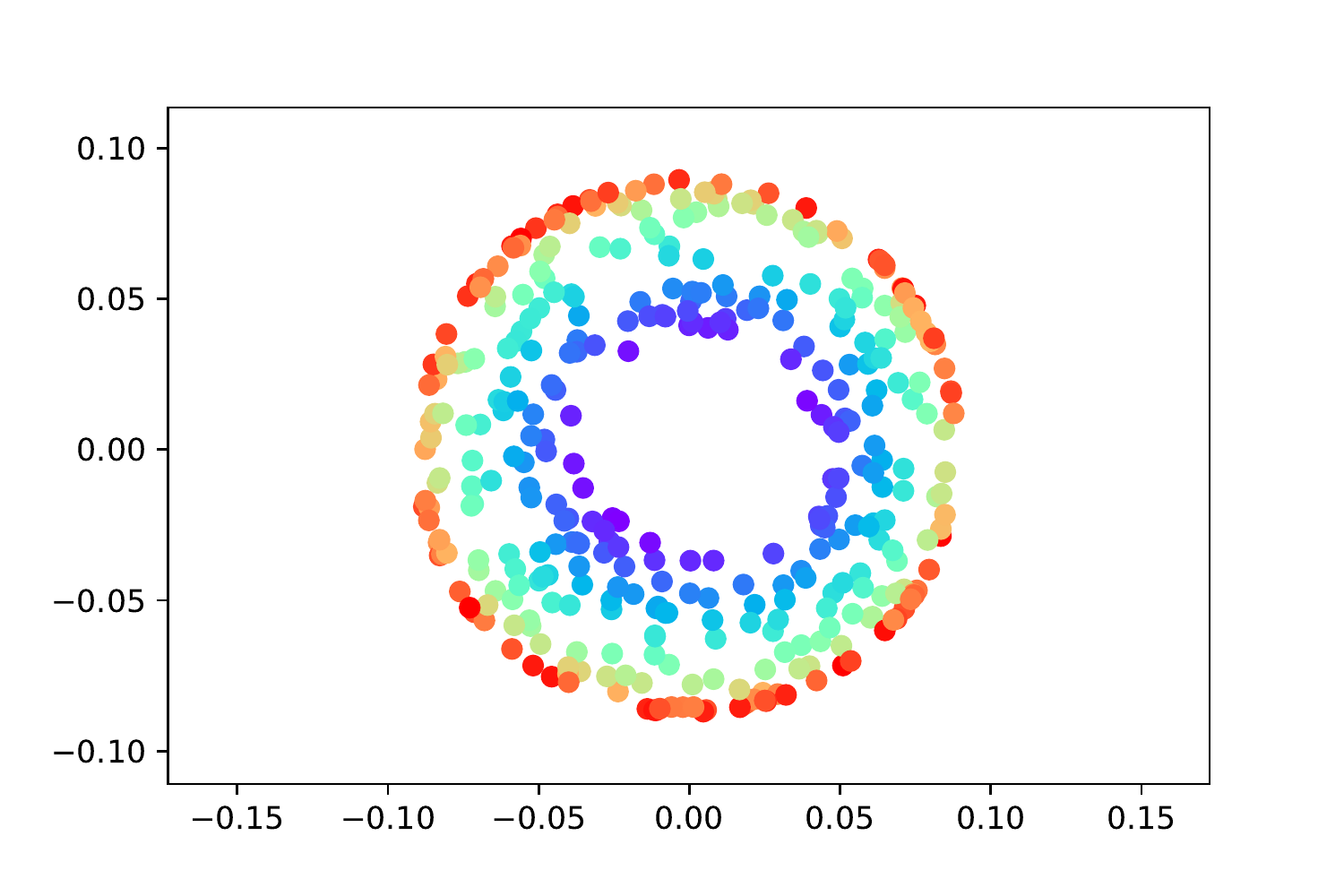}}
    \caption{The illustration of locally linear embedding. (a) A two-dimensional manifold; (b) the three-dimensional data points sampled from (a), colored according to the $z$-coordinates; (c) the data points after dimensionality reduction by LLE.}
    \label{fig:lle_illustration}
\end{figure}

Locally linear embedding (LLE), proposed by Roweis and Saul \cite{roweis2000}, is an unsupervised learning algorithm that computes low-dimensional, topology-preserving embeddings of high-dimensional data points. LLE is an instance of kernel principal component analysis (kernel PCA), which handles nonlinear dimensionality reduction \cite{scholkopf1998}. As illustrated in Fig.~\ref{fig:lle_illustration}, LLE maps high-dimensional data into a single global coordinate system of lower dimensionality.

In this paper, we use LLE to accelerate the computation of the phase field. The main idea is that from the offline calculation of enough cracked microstructures, we will be able to reconstruct crack evolution due to various initial crack patterns with minimal computation online. 

The specific process of LLE is as follows. Suppose that there are $N$ input data points $\bm{X}_i\in \mathbb{R}^{\mathcal{D}}$ where $ i = 1,  ..., N$, each $\bm{X}_i$ containing the phase field values representing a specific cracked microstructure. According to \cite{roweis2000}, under the assumption that all inputs are on the same manifold, we can linearly reconstruct each data point $\bm{X}_i$ by its $k_1$ ($\ll N$) nearest neighbors, say
\begin{equation}
    \bm{X}_i = \sum_{j\in S_i} W_{ij}\bm{X}_{j}, \label{eq:linear_recon}
\end{equation}
where $W_{ij}$ are the weights to be determined and $S_i$ represents the set of the $k_1$ nearest neighbors of $\bm{X}_i$ in the $l^2$-norm. 

To compute these weights $W_{ij}$, we minimize the cost function which measures the reconstruction errors:
\begin{equation}
    \mathcal{F}(\bm{W}) = \sum_{i=1}^N \left\lVert \bm{X}_i - \sum_{j\in S_{i}} W_{ij}\bm{X}_{j}\right\rVert^2. \label{eq:cost_func}
\end{equation}
The minimization of $\mathcal{F}(\bm{W})$ is subjected to two constraints: (i) each data point $\bm{X}_i$ is reconstructed only from its neighbors: $W_{ij}=0$ if $\bm{X}_j\notin S_{i}$. (ii) the rows of the weight matrix sum to 1: $\sum_{j\in S_{i}} W_{ij} = 1$, $ i = 1,  ..., N$. An important feature is, for any data point, the weights are invariant to rotation, rescaling and translation of that data point with respect to its neighbors \cite{roweis2000}.

Now we suppose that all data points are mapped into a lower dimensional embedding space (manifold) of dimension $\mathcal{L}$, $\mathcal{L}\ll\mathcal{D}$. The reconstruction weights $W_{ij}$ remain unchanged in such transformation. Therefore, each high dimensional data point $\bm{X}_i$ is mapped to a low dimensional vector $\bm{Y}_i$ representing coordinates on the manifold. We compute $\bm{Y}:=\{\bm{Y}_i\}$ by minimizing the embedding cost function
\begin{equation}
    \mathcal{G}(\bm{Y}) = \sum_{i=1}^N \left\lVert \bm{Y}_i - \sum_{j\in S_{i}} W_{mi}\bm{Y}_{j}\right\rVert^2, \quad \bm{Y} := \{\bm{Y}_i\}. \label{eq:embedding_cost_func}
\end{equation}
During this minimization, the weights $W_{ij}$ are fixed. To fully determine $\{\bm{Y}_i\}$, certain constraints have to be imposed so that the solution is unique \cite{roweis2000}. The resulting constrained minimization problem can be solved via an $N\times N$ eigenvalue problem.


\subsection{Training and Output Reconstruction}\label{sec:train}
As previously discussed, the offline procedure of this manifold learning scheme consists of two stages: (1) dataset generation with the phase field analysis for the RVE, (2) data manifold construction with LLE. Then for any given phase field under the same load, the online reconstruction procedure readily delivers the phase field evolution. 

\begin{figure}[htpb]
    \centering
    \subfigure[]{\includegraphics[width=0.46\columnwidth]{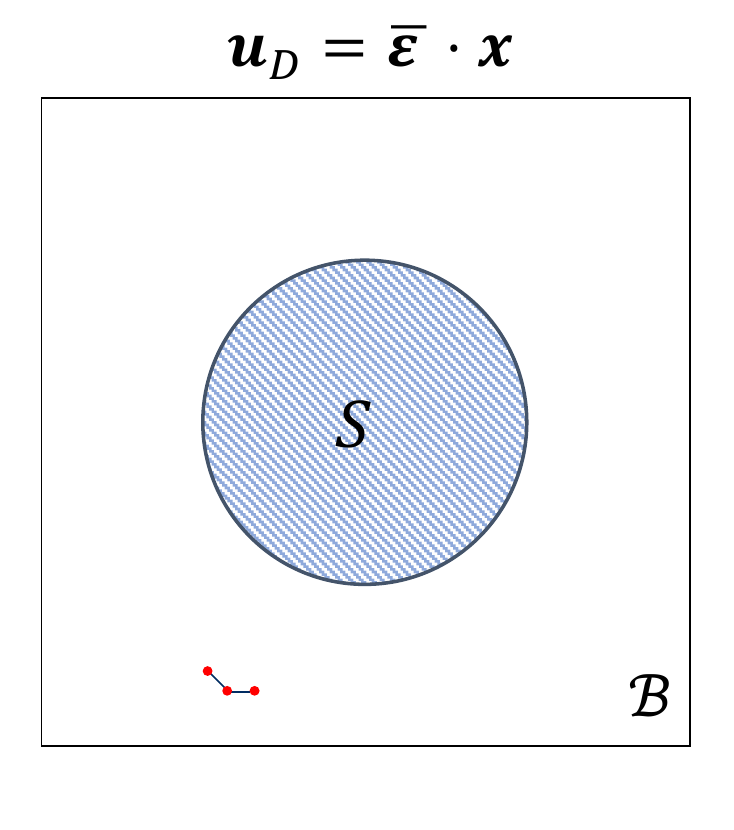}}
    \subfigure[]{\includegraphics[width=0.48\columnwidth]{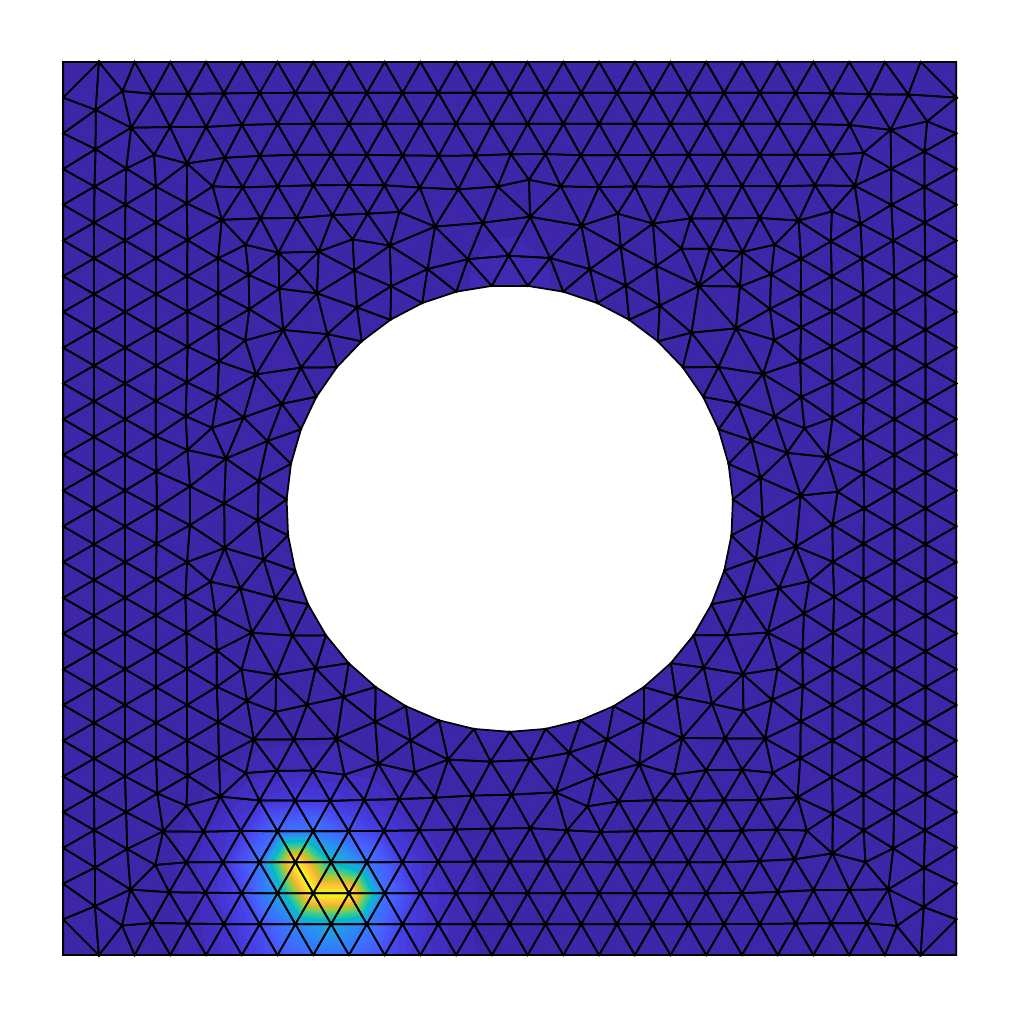}}
    \caption{(a) Setup of the boundary value problem for the RVE; (b) mesh and a typical initial phase field.}
    \label{fig:config}
\end{figure}

To generate the training data, we subject a series of RVEs with an initial crack at various locations to the unilateral tension test.  The configuration and mesh with an initial phase field are shown in Fig.~\ref{fig:config}. The mesh shown in Fig.~\ref{fig:config}(b) contains $\mathcal{D}$ nodes, so every input data point $\bm{X}_i$ as well as the corresponding output data point $\bm{Z}_i$ is a column vector with $\mathcal{D}$ phase field values. 

Here we made some simplifications for the micro crack simulation so that we can better illustrate the main idea: (1) as mentioned in Section \ref{fe2}, the load is a unilateral tension with given displacement as shown in \eqref{eq:boundary}, where the macroscopic strain is $\overline{\bm{\varepsilon}} = \overline{\varepsilon}_{22} \bm{e}_2 \otimes \bm{e}_2$; (2) we only consider cracks in the matrix and ignore those on the interface and in the fiber; (3) the initial crack consists of two edges and three connected nodes, but nodes belonging to the same element are forbidden.

\begin{figure}[htpb]
	\centering
	\includegraphics[width=\columnwidth]{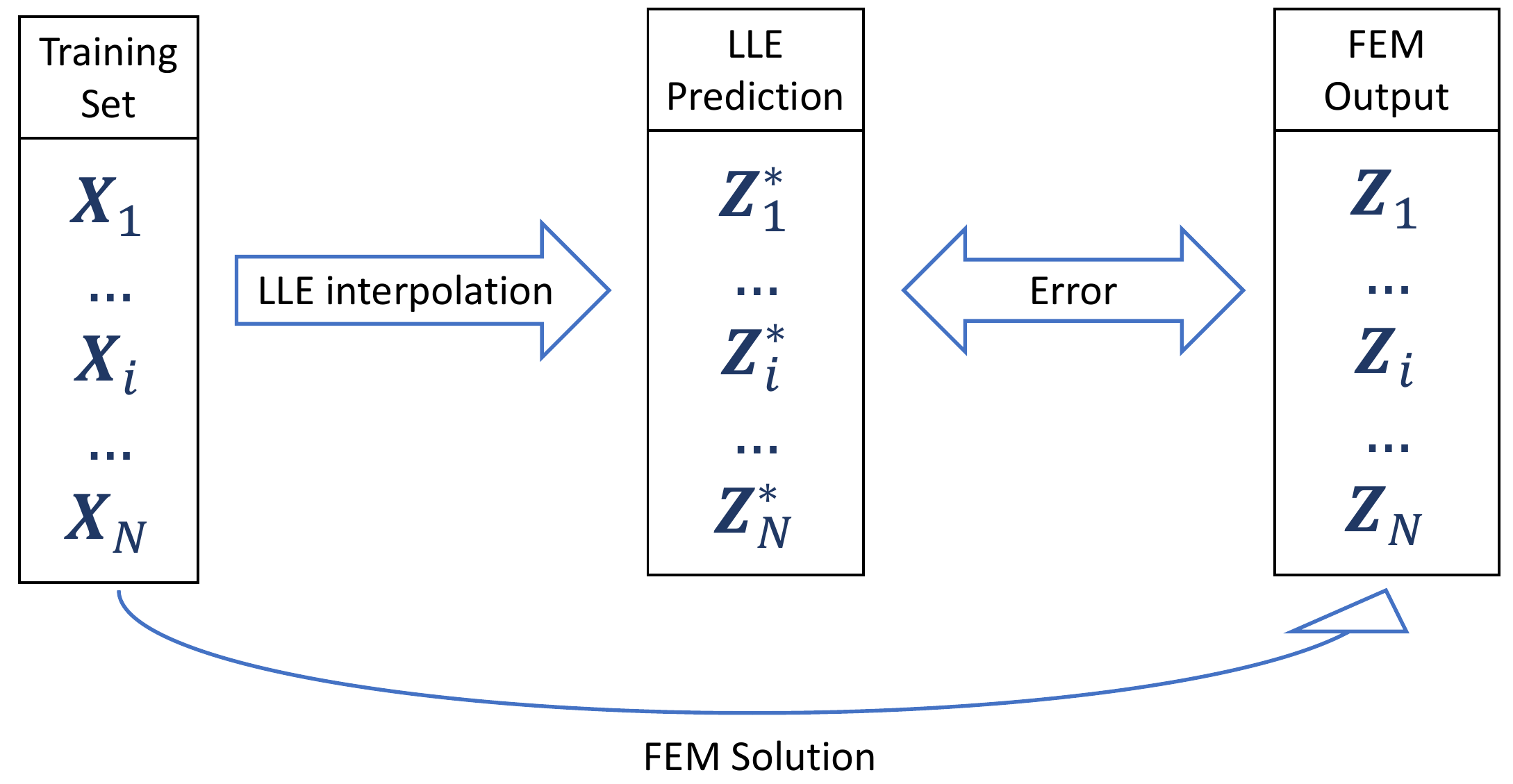}
	\caption{The process of manifold learning using LLE.}
	\label{fig:lle_process}
\end{figure}

With the phase field values $d=1$ imposed at the three nodes mentioned in (3) above and with an all-zero displacement field $\bm{u}\equiv\bm{0}$, we minimize \eqref{eq:variational} to get an ``equilibrated'' phase field as a typical input $\bm{X}_i$. The totality of such inputs is termed the \emph{training set}.
The process of construction of the data manifold with the training set is illustrated in Fig.~\ref{fig:lle_process}.  

For each input $\bm{X}_i$ in the training set, we generate the high-fidelity solution of the evolved phase field through a finite element program, and the result is denoted $\bm{Z}_i$. Notice that only input data are used during the LLE construction, while the output data $\left\{\bm{Z}_i\right\}$ are only used for reconstruction. The output data are not limited to be the phase field solution at the given load, nor need it have the same dimension as the input datapoints.

Once we obtain the data manifold, we reconstruct the output, marked by $\bm{Z}^{*}_i$, for every new input $\bm{X}^{*}_i$ \emph{not} in the training set through the following process: 
\begin{enumerate}
    \item We find $k_2$ ($\ll N$, which can be the same as $k_1$, see Section \ref{numerical} for more details) nearest neighbors of $\bm{X}_i^*$ in $\bm{X}$ and the corresponding weights in the high dimensional space $\mathbb{R}^\mathcal{D}$, then we map $\bm{X}_i^*$ to the low dimensional manifold $\bm{Y}_i^*\in\mathbb{R}^\mathcal{L}$.
    \item We find the $k_2$ nearest neighbors of $\bm{Y}_i^*$ in $\bm{Y}$, called $S_i^*$, and their weights $W_{ij}$ in the low dimensional manifold. Note that these neighbors may not correspond to those in the previous step. 
    \item Locally linear reconstruct the output with weights and its $k_2$ nearest neighbors in high dimensional data space: \[\bm{Z}_i^{*}=\sum_{j\in S_i^*}W_{ij}\bm{Z}_j.\]
\end{enumerate}

\section{Numerical Implementation and Validation}
\label{numerical}

In this section, we detail the numerical implementation and results of the proposed manifold learning method. 
In addition, we provide a validation check for the computational strategy.


\subsection{Data Generation}
In our high-fidelity finite element analysis, the material constants are chosen as according to Table \ref{tab:material}. The RVE size $L=500$mm and the macroscopic strain  $\overline{\bm{\varepsilon}} = \overline{\varepsilon}_{22} \bm{e}_2 \otimes \bm{e}_2$ where $\overline{\varepsilon}_{22}=1.4\times10^{-4}$. The regularized length scale parameter $l$ is chosen such that $h\le l/2$, where $h$ is the mesh size. We randomly generated 496 initial phase fields as detailed in Section \ref{sec:train}, which correspond to 496 data points for training (manifold learning).
\begin{table}[htpb]
    \scriptsize
    \caption{Material parameters used in the high-fidelity finite element simulations}
    \label{tab:material} 
    \begin{tabular}{cccccc}
        \hline\noalign{\smallskip}
        $\lambda$(GPa) & $\mu$(GPa) & $\lambda_1$(GPa) & $\mu_1$(GPa) & $g_c\mathrm{(mJ/mm^2)}$ & $l$(mm)  \\
        \noalign{\smallskip}\hline\noalign{\smallskip}
        121.15 & 80.77 & 105.58 & 172.27 & 2.7 & 40\\
        \noalign{\smallskip}\hline
    \end{tabular}
\end{table}

\subsection{Parameter Selection by Cross Validation}
Once the data points are generated, parameter selection is conducted for the manifold learning and reconstruction. Recall the LLE manifold is defined by two hyperparameters $k_1$ and $\mathcal{L}$, 
while the reconstruction process is defined by one hyperparameter $k_2$. Hence\added{,} the complete manifold model for the problem requires three hyperparameters ($k_1, k_2, \mathcal{L}$).

\begin{figure}[htpb]
    \centering
    \includegraphics[width=1\columnwidth]{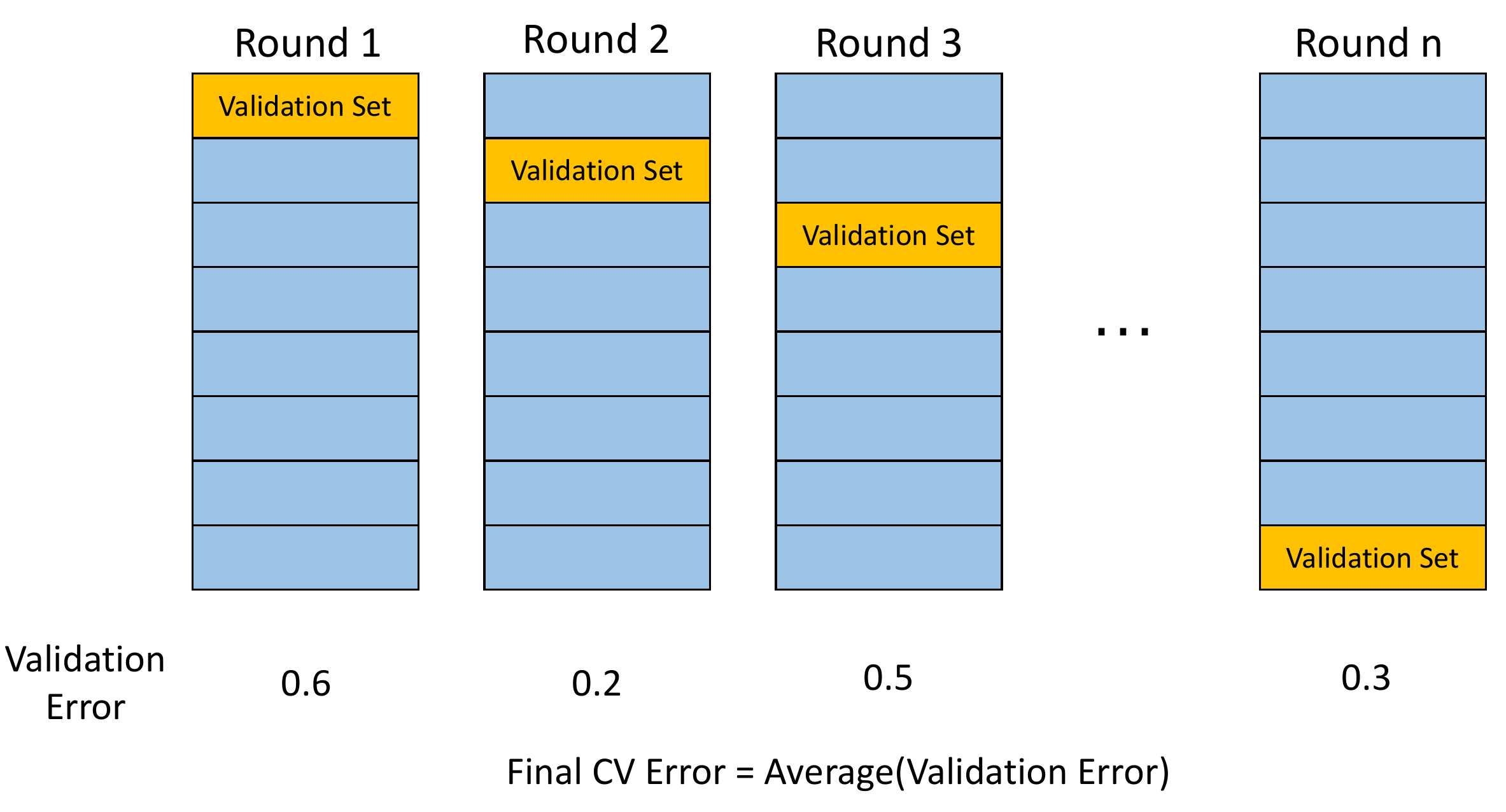}
    \caption{The process of cross validation.}
    \label{fig:cv}
\end{figure}

The adopted parameter selection method is called cross validation (CV).  Through CV we will select the best combination of hyperparameters which leads to a balance of cost and accuracy. 
The CV process is proceeded as follows. We split the whole dataset ($N=490$ datapoints) to be $n=10$ equal-sized mutually disjoint subsets randomly, $\bm{X}^{(1)}$,\ldots,$\bm{X}^{(n)}$, then we choose $n-1$ subsets as the training set to generate the manifold, and use the remaining one for validation, say the $j$th subset $\bm{X}^{(j)}$. 
Let $\bm{Z}^{(j)}=\{\bm{Z}_i^{(j)}\}$ denote the corresponding output phase field data for the validation set, and $\bm{Z}^{*(j)}=\{\bm{Z}^{*(j)}_i\}$ the LLE reconstruction. Then the final CV error $R$ reads
\[
R = \frac{1}{n} \sum_{j=1}^{n} \sum_i \frac{\lVert\bm{Z}_i^{*(j)}-\bm{Z}_i^{(j)}\rVert_{l^2}}{\lVert\bm{Z}_i^{(j)}\rVert_{l^2}}.
\]
This procedure is illustrated in Fig.~\ref{fig:cv}.

The procedure to select hyperparameters consists of two stages: (1) the dimension reduction process involving $k_1$, and (2) the reconstruction process involving $k_2$.  Iterating through combinations of $(\mathcal{L}, k_1,k_2)$ with a fixed $k_2$ value, an error matrix is deduced with columns denoting values of $k_1/k_2$, and rows denoting values of $\mathcal{L}$ as shown in Table \ref{tab:cv_error1}. 
We find that $k_1=k_2$ will yield a low CV error, which is reasonable, as the case $k_1>k_2$ will lead to information loss in the reconstruction process, and $k_1<k_2$ will add noise to the reconstruction process. 

\begin{table}[htpb]
    \scriptsize
    \caption{CV error with different combinations of $k_1/k_2$ and $\mathcal{L}$, with $k_2=20$.}
    \label{tab:cv_error1} 
    \begin{tabular}{ccccccc}
        \hline\noalign{\smallskip}
        $k_1/k_2\backslash \mathcal{L}$ & 20 & 40 & 60 & 80 & 100 & 120  \\
        \noalign{\smallskip}\hline\noalign{\smallskip}
        1/4 & 0.4798 & 0.4042 & 0.3769 & 0.3599 & 0.3529 & 0.3442\\
        \noalign{\smallskip}\hline\noalign{\smallskip}
        1/2 & 0.4021 & 0.3645 & 0.3467 & 0.3338 & 0.3261 & 0.3222 \\
        \noalign{\smallskip}\hline\noalign{\smallskip}
        1 & 0.3754 & 0.3472 & 0.3260 & 0.3114 & 0.3059 & 0.3036\\
        \noalign{\smallskip}\hline\noalign{\smallskip}
        2 & 0.3768 & 0.3472 & 0.3254 & 0.3103 & 0.3024 & 0.3007\\
        \noalign{\smallskip}\hline\noalign{\smallskip}
        4 & 0.3774 & 0.3457 & 0.3241 & 0.3096 & 0.3026 & 0.3001\\
        \noalign{\smallskip}\hline
    \end{tabular}
\end{table}

Then we fix $k_1=k_2$ and perform more CV to obtain Table \ref{tab:cv_error2},
from which we determine that $k_1=k_2=20$ gives a relatively low CV error for each $\mathcal{L}$.

\begin{table}[htpb]
    \scriptsize
    \caption{CV error with different combinations of $k_1 (= k_2)$ and $\mathcal{L}$.}
    \label{tab:cv_error2} 
    \begin{tabular}{ccccccc}
        \hline\noalign{\smallskip}
        $k_1\backslash \mathcal{L}$ & 20 & 40 & 60 & 80 & 100 & 120  \\
        \noalign{\smallskip}\hline\noalign{\smallskip}
        5 & 0.6293 & 0.4905 & 0.4452 & 0.4229 & 0.4015 & 0.3850 \\
        \noalign{\smallskip}\hline\noalign{\smallskip}
        10 & 0.4109 & 0.3665 & 0.3461 & 0.3343 & 0.3275 & 0.3281 \\
        \noalign{\smallskip}\hline\noalign{\smallskip}
        15 & 0.3715 & 0.3432 & 0.3244 & 0.3147 & 0.3082 & 0.3061\\
        \noalign{\smallskip}\hline\noalign{\smallskip}
        20 & 0.3754 & 0.3472 & 0.3260 & 0.3114 & 0.3059 & 0.3036\\
        \noalign{\smallskip}\hline\noalign{\smallskip}
        25 & 0.3914 & 0.3582 & 0.3323 & 0.3148 & 0.3070 & 0.3036\\
        \noalign{\smallskip}\hline\noalign{\smallskip}
        30 & 0.4073 & 0.3694 & 0.3383 & 0.3175 & 0.3081 & 0.3046\\
        \noalign{\smallskip}\hline
    \end{tabular}
\end{table}

\begin{figure}[htpb]
    \centering
    \includegraphics[width=\columnwidth]{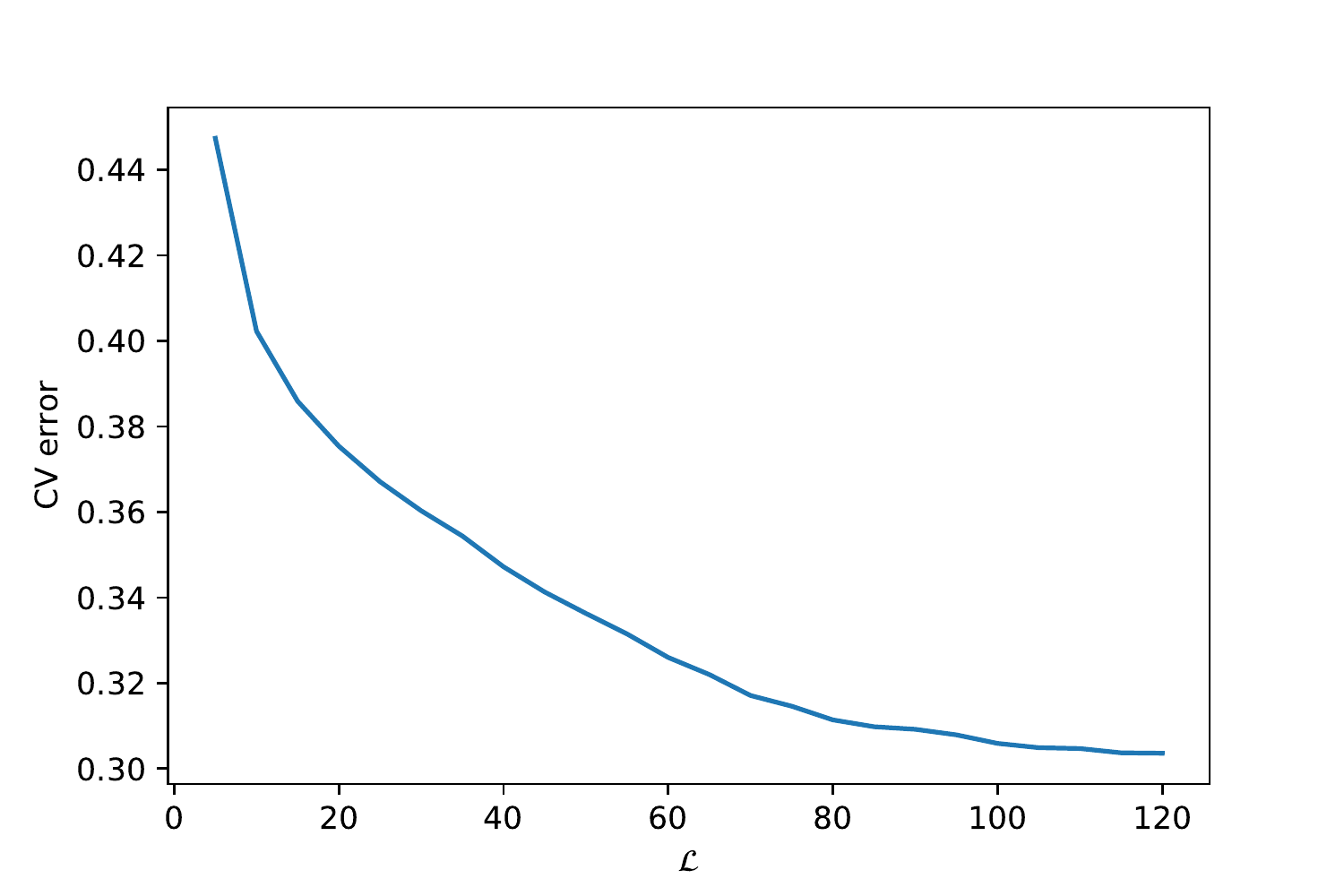}
    \caption{CV error vs.~$\mathcal{L}$.}
    \label{fig:cv_errors}
\end{figure}

Then, we plot the CV error as a function of $\mathcal{L}$ in Fig.~\ref{fig:cv_errors}.
This figure indicates that an increase in $\mathcal{L}$ will reduce the average error, as expected. However, using a larger $\mathcal{L}$ increases the training time. Therefore, we follow the standard way to make the trade-off, i.e., to get the critical turning point at approximately the elbow, where $\mathcal{L}=80$. When $\mathcal{L}$ is beyond this value, the error decreases at a very slow rate, while the training efficiency continually decreases.

In conclusion, the chosen hyperparameters are $(k_1,k_2,\mathcal{L})=(20,20,80)$.


\subsection{Results and Discussion}
\begin{figure}[htpb]
    \centering
    \includegraphics[width=\columnwidth]{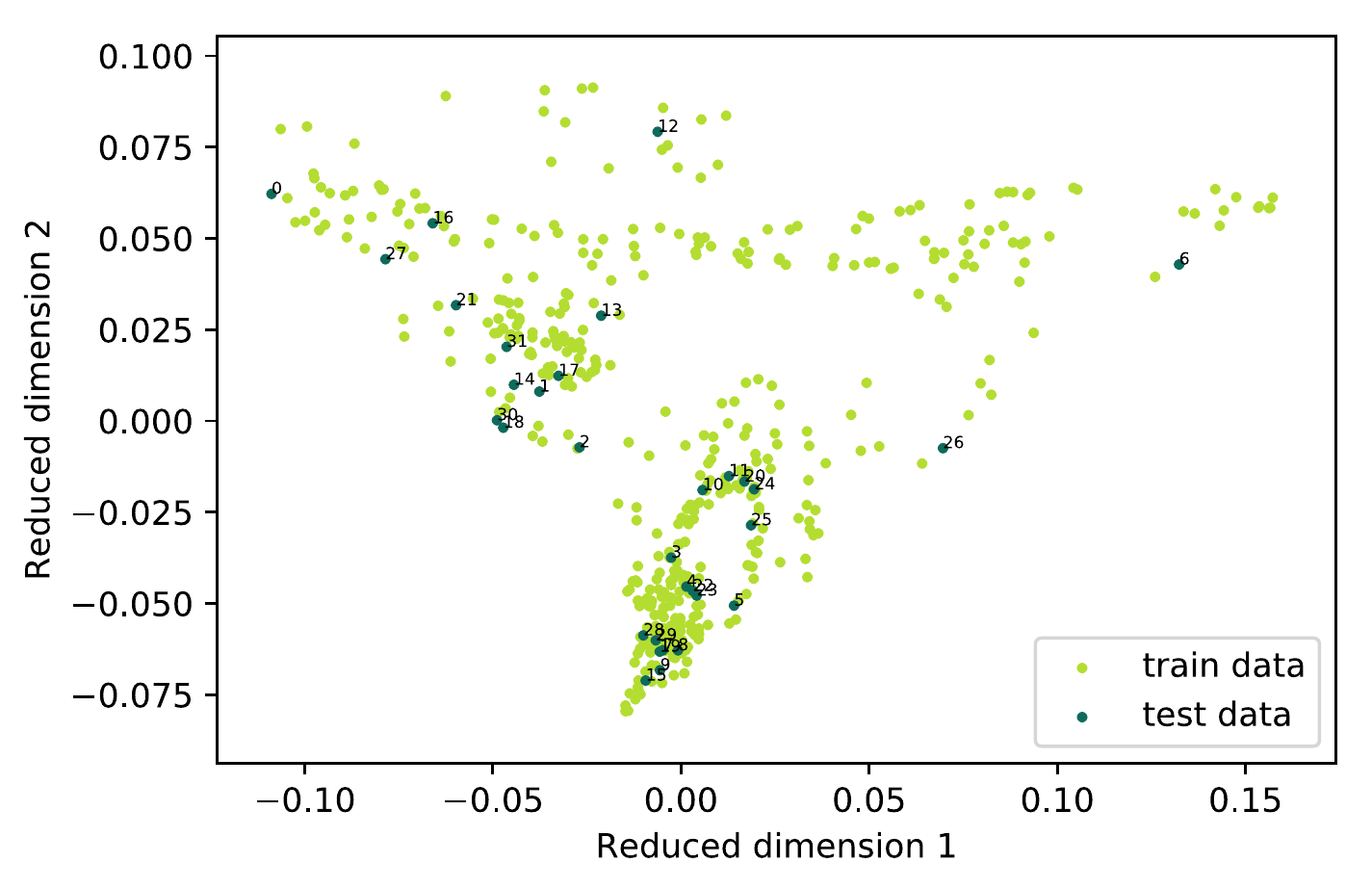}
    \caption{2D visualization of the 80D manifold}
    \label{fig:2d_70d}
\end{figure}

To remove the data bias, we generate a new set of 496 data points, which shares no data points with the set used in the parameter selection process. With the selected hyperparameters $(k_1,k_2,\mathcal{L})=(20,20,80)$, we build the model using 464 data points for training, and use the remaining 32 data points for testing. To visualize the manifold built by the training data, and together showing the test data, we perform an LLE reduction again for the 80-dimensional manifold to 2 dimensions, as in Fig.~\ref{fig:2d_70d}. It can be observed that the test data points are not far from the manifold trained from the training data.

\begin{figure}[htpb]
    \centering
    \subfigure[]{\includegraphics[width=\columnwidth]{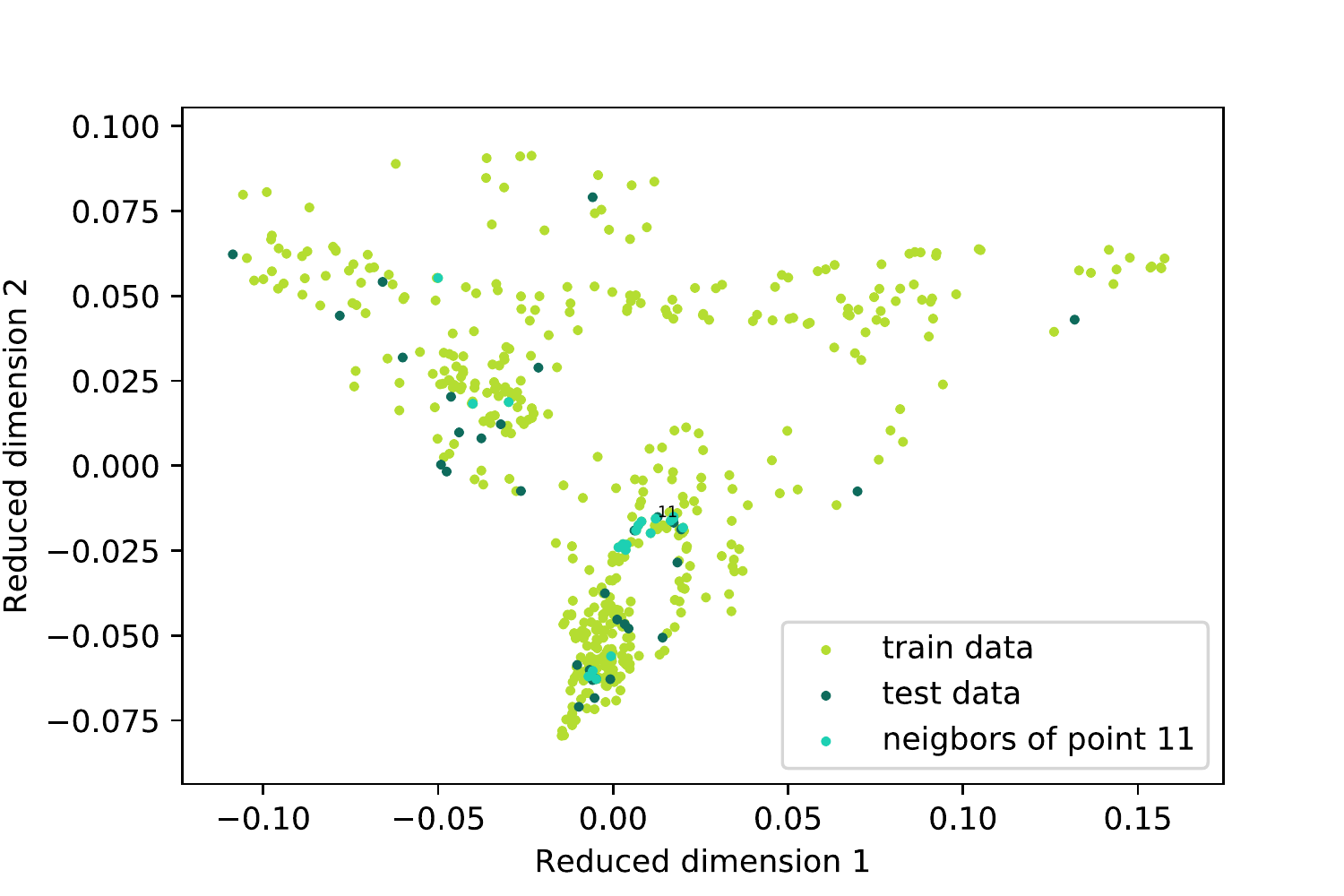}}
    \subfigure[]{\includegraphics[width=\columnwidth]{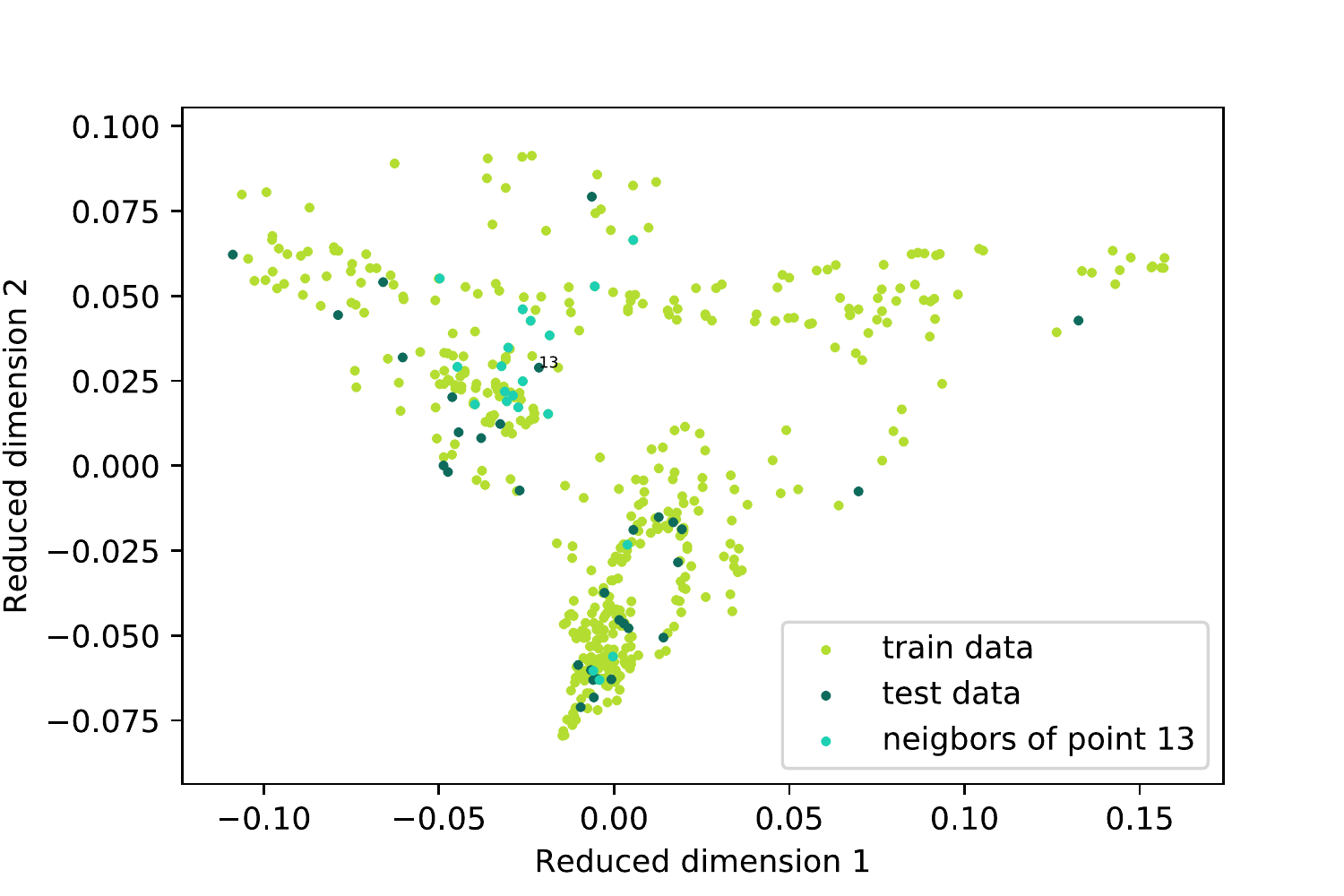}}
    \caption{(a) Nearest 20 neighbor points of test point No.~11; (b) Nearest 20 neighbor points of test point No.~13.}
    \label{fig:2d_group}
\end{figure}


Next we extract and visualize the nearest neighbors of a certain data point, as shown in Fig.~\ref{fig:2d_group}(a) and (b).

In Fig.~\ref{fig:2d_group}(a), we observe that the nearest neighbors in the training set are close to the chosen test data point (Point No.~11). In Fig.~\ref{fig:2d_group}(b), however, the nearest neighbors of the chosen test data point (Point No.~13) appear scattering around. 
This phenomenon is still acceptable since the distances between points in the remaining 78 dimensions are not seen in the figures. 

\begin{figure*}[htpb]
    \centering
    \includegraphics[width=2\columnwidth]{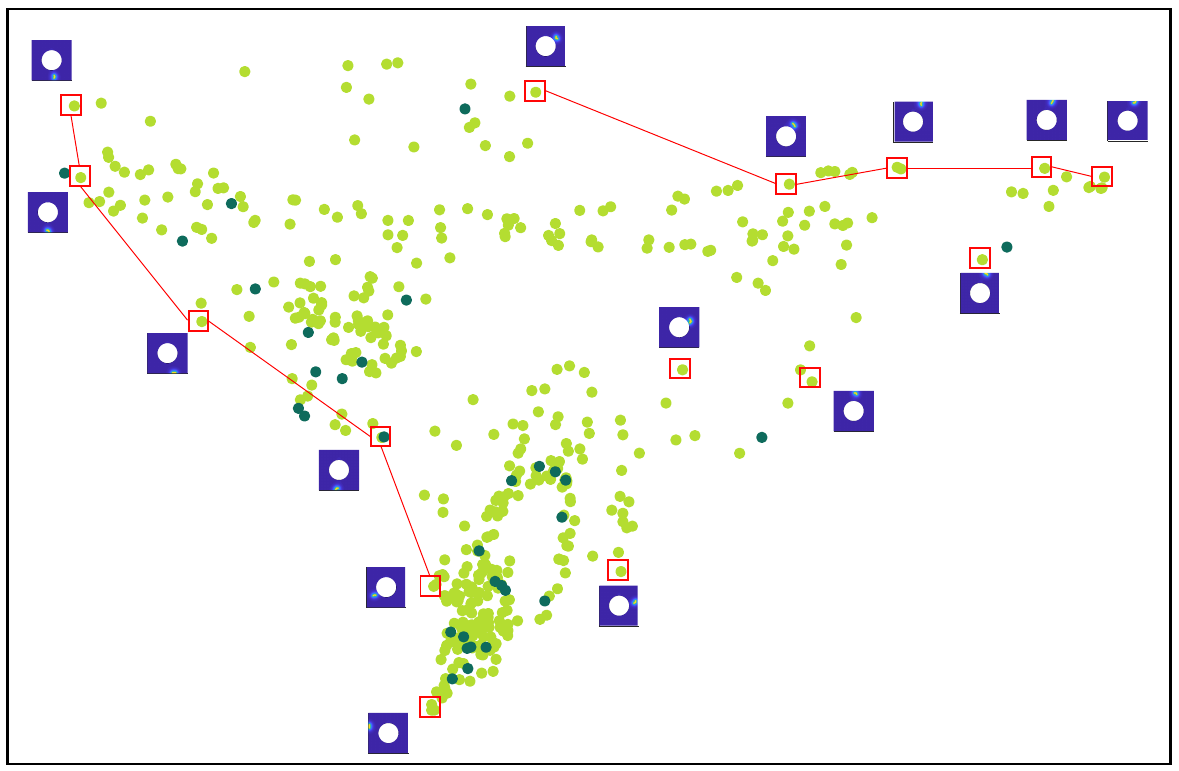}
    \caption{Crack microstructures mapped into the manifold described by the first two coordinates of LLE. Representative microstructures are shown next to the square points. The solid line shows a continuous mode change of microstructures.}
    \label{fig:cluster}
\end{figure*}

We next visualize the cracked microstructures in Fig.~\ref{fig:cluster}, where we can observe a pattern that similar microstructure will cluster in a continuous mode, showing the dimension reduction is reasonable.

\subsection{Reconstruction Error Analysis for the Phase Field}
\begin{figure}[htpb]
    \centering
    \includegraphics[width=\columnwidth]{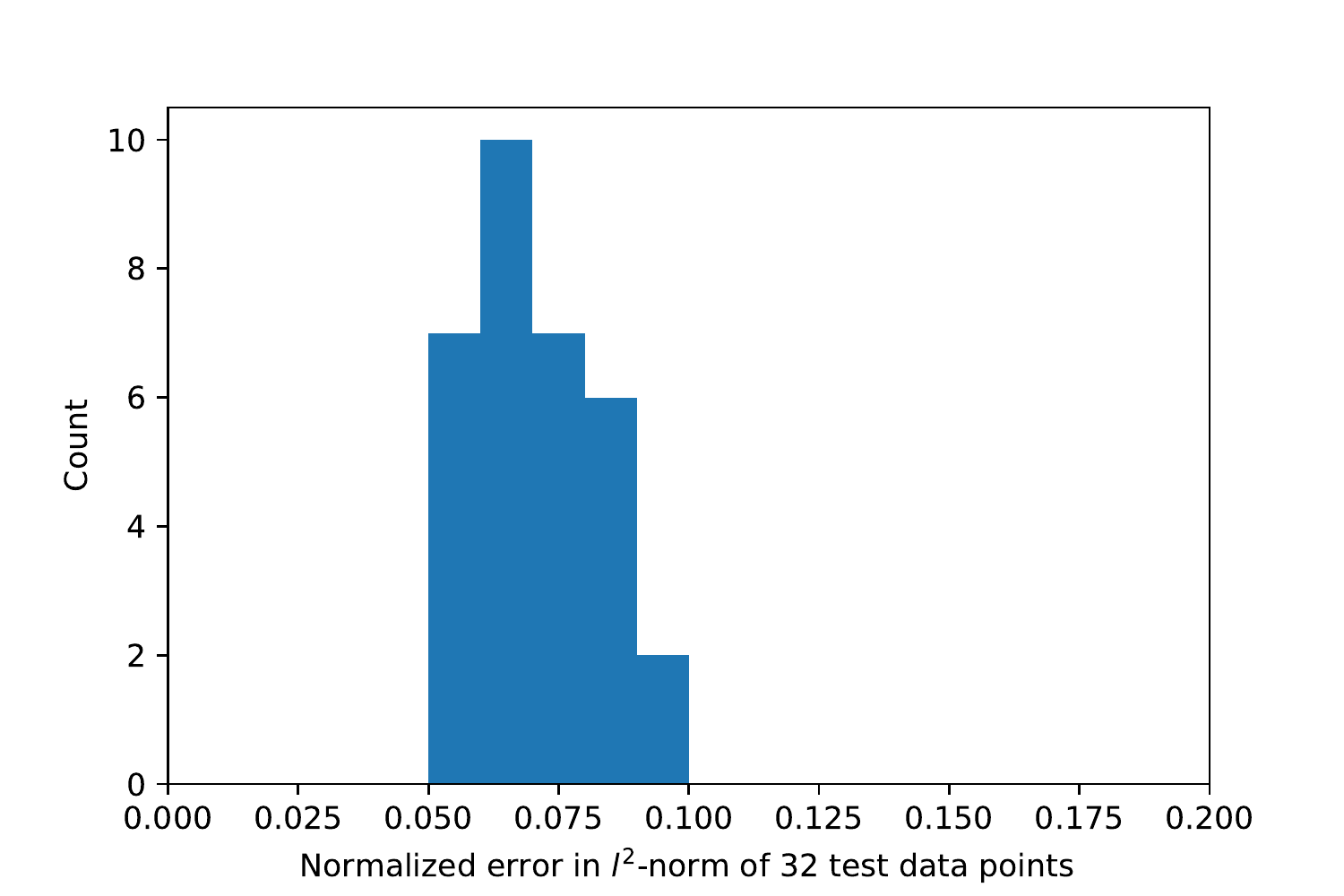}
    \caption{Normalized $l^2$ reconstruction error of the evolved phase field, i.e., $\lVert \bm{Z}^*_i - \bm{Z}_i\rVert_{l^2}/\lVert \bm{Z}_i \rVert_{l^2}$, of the 32 test data points.}
    \label{fig:absolute_error}
\end{figure}

In this subsection, the output is the evolved phase field $\bm{Z}^*_i=\left\{d_{j}\right\}_i$, where $j=1,2,...,\mathcal{D}$. Therefore, the output $\bm{Z}^*_i$ and input $\bm{X}_i$ have the same dimension.
A histogram showing the reconstruction errors is given in Fig.~\ref{fig:absolute_error}, where we use the normalized $l^2$-norm to represent the error magnitude in the output phase field, i.e., 
\begin{equation}
    \label{eq:recon_error}
    \frac{\lVert \bm{Z}^*_i - \bm{Z}_i\rVert_{l^2}}{\lVert \bm{Z}_i \rVert_{l^2}}.
\end{equation}
From this figure it can be seen that the LLE reconstruction error for the phase field is acceptable. 

\begin{figure}[htpb]
    \centering
    \includegraphics[width=\columnwidth]{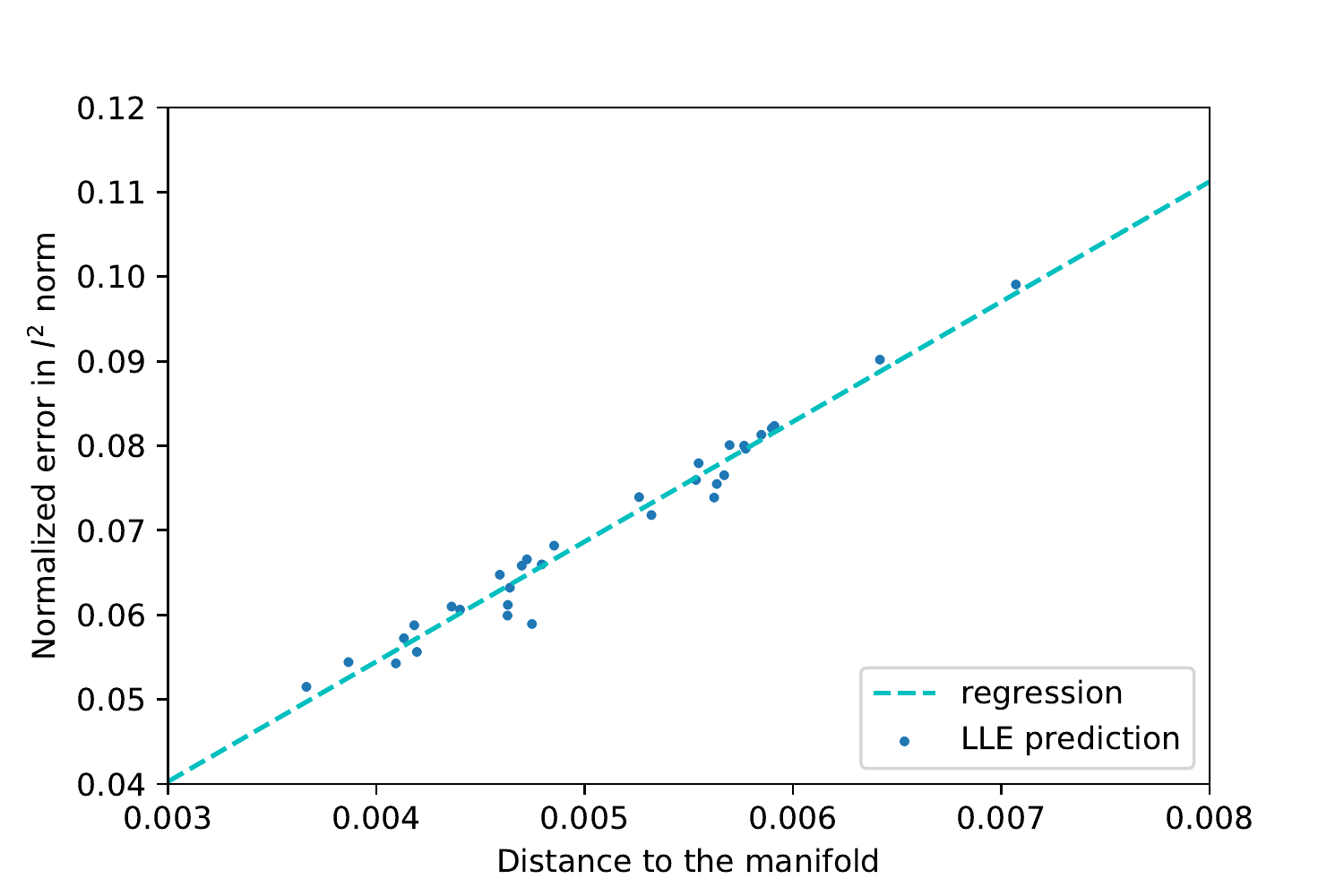}
    \caption{Normalized $l^2$ error of the evolved phase field vs.~the distance to the manifold. The $l^2$ errors have a positive correlation with the distance to the manifold.}
    \label{fig:distance_error}
\end{figure}

To examine the deciding factor of such error, we plot the normalized error in $l^2$-norm of the 32 test points versus their distance to the manifold in
Fig.~\ref{fig:distance_error}. Here the distance of $\bm{X}_i^*$ to the manifold is given by \[\left\|\bm{X}_i^*-\sum_{j\in S_i^*} W_{ij} \bm{X}_j\right\|_{l^2}.\] A positive relationship between the reconstruction error and this distance is observed, without outliers. Thus we can safely say that if a test data point is close enough to the manifold, the reconstruction error of its microcrack propagation result will be small, guaranteeing the validity of this LLE manifold learning method. 

\subsection{Reconstruction Error Analysis for the Homogenized Stress}
\begin{figure}[htpb]
    \centering
    \includegraphics[width=\columnwidth]{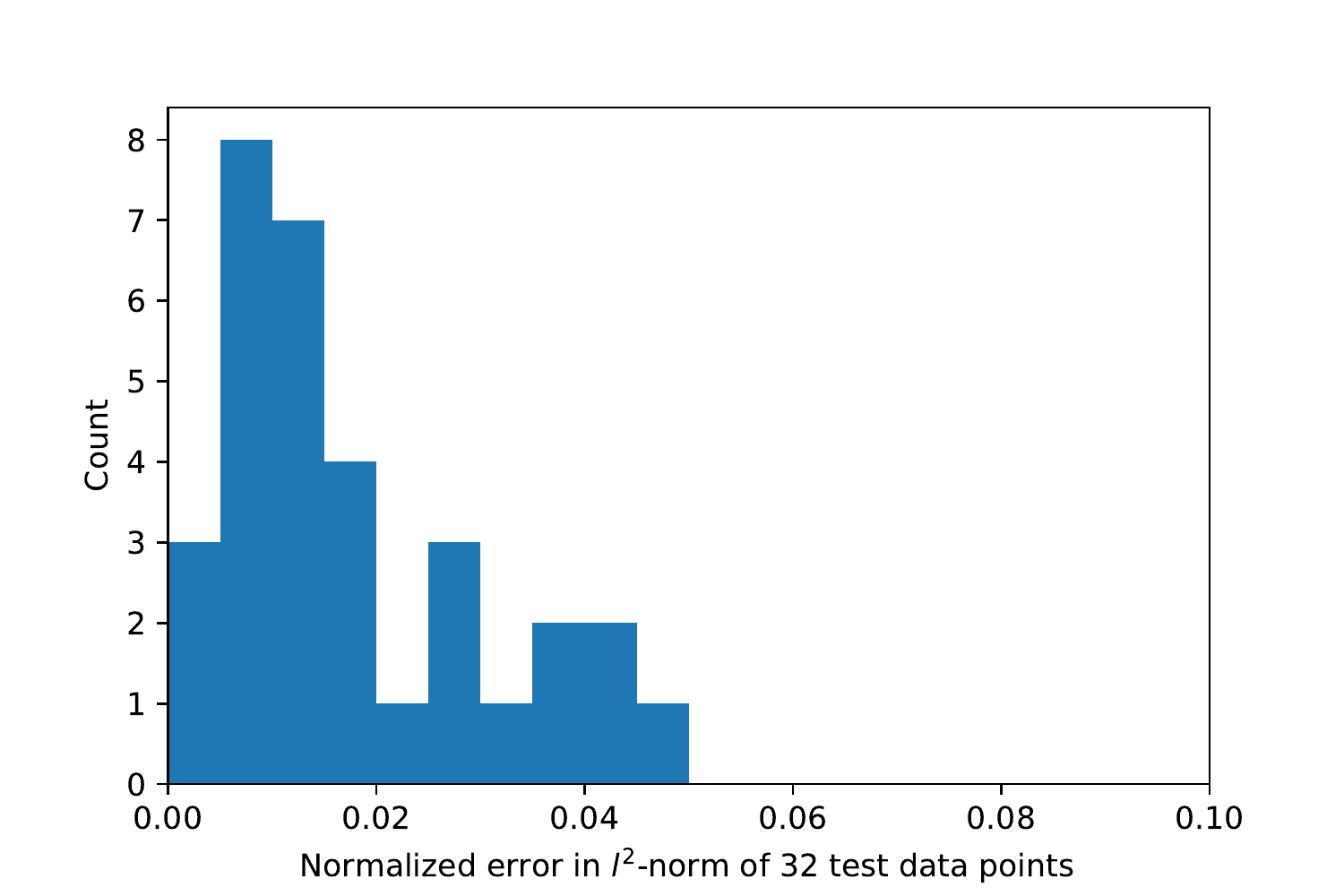}
    \caption{Normalized $l^2$ reconstruction error of the homogenized stress, i.e., $\lVert \overline{\bm{\sigma}}^*_i - \overline{\bm{\sigma}}_i\rVert_{l^2}/\lVert \overline{\bm{\sigma}}_i \rVert_{l^2}$, of the 32 test data points.}
    \label{fig:recon_error_stress}
\end{figure}

\begin{figure}[htpb]
    \centering
    \includegraphics[width=\columnwidth]{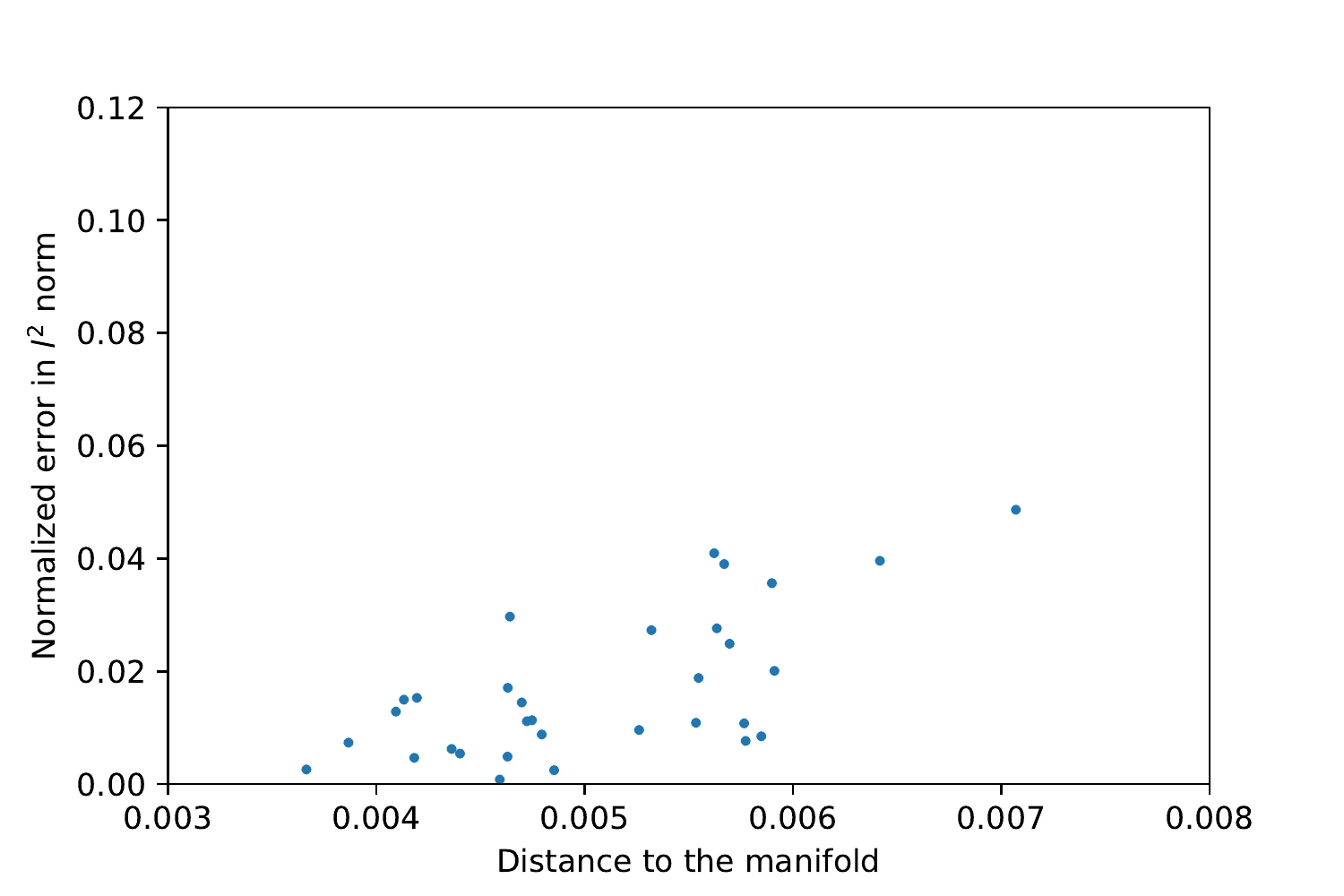}
    \caption{Normalized $l^2$ error of the homogenized stress vs.~the distance to the manifold. The $l^2$ errors are bounded by a factor times the distance to the manifold.}
    \label{fig:distance_error_stress}
\end{figure}

In this subsection, the output is the homogenized stress $\bm{Z}^*_i=\overline{\bm{\sigma}}_i$, where $\overline{\bm{\sigma}}_i=\left\{\overline{\sigma}_x, \overline{\sigma}_y, \overline{\sigma}_z, \overline{\sigma}_{xy}\right\}_i$, where for the plane strain case, $\overline{\sigma}_z=\nu(\overline{\sigma}_x+ \overline{\sigma}_y)$ for the matrix and likewise for the fiber. As Fig.~\ref{fig:fe2scheme} shows, the homogenized stress is obtained from the RVE through the volume average,
\begin{equation*}
    \overline{\bm{\sigma}}=\frac{1}{|\mathcal{B}|}\int_{\partial \mathcal{B}} \bm{\sigma}\; \mathrm{d}\mathcal{B}.
\end{equation*}
Then the normalized reconstruction error in $l^2$-norm \eqref{eq:recon_error}  becomes
\begin{equation}
    \frac{\lVert \overline{\bm{\sigma}}^*_i - \overline{\bm{\sigma}}_i\rVert_{l^2}}{\lVert \overline{\bm{\sigma}}_i \rVert_{l^2}}.
\end{equation}
The normalized reconstruction error of the homogenized stress is shown in Fig.~\ref{fig:recon_error_stress}. It shows that the normalized reconstruction error is smaller than $0.05$, which is very small. 
Fig.~\ref{fig:distance_error_stress} shows that the reconstruction error is bounded by a factor times the distance to the manifold, indicating a similar conclusion, i.e., an \emph{a priori} error estimate can be obtained.

\paragraph{Remark.} Through the correlation of the reconstruction error and the distance to the manifold, we can pre-determine whether a new input data point $\bm{X}_i^*$ is suitable for the manifold learning approach: if $\bm{X}_i^*$ is close enough to the manifold, the reconstruction of the phase field at the given load will be accurate; otherwise, if it is far away from the manifold, we should either not use the manifold reconstruction for this particular input, or augment the training set with $\bm{X}_i^*$. This property can also be exploited to aid an adaptivity procedure to augment the training set on the fly: if the distance from a certain new input $\bm{X}_i^*$ to its manifold projection is too high, then we can add it (and its output from high-fidelity computation) to the training set.




\section{Conclusions}
\label{conclusions}

We have proposed a manifold learning approach to accelerate phase field fracture simulations in the RVE in the context of the FE$^2$ scheme. Considering a group of RVEs with the same microstructure except for the microcracks, we use the phase field approach to represent such microcracks. 
    We then make use of the LLE technique to construct a data manifold that contains a collection of similar cracked microstructures (RVEs).
This LLE manifold can be used to efficiently and accurately predict the phase field output as a function of the initial phase field, provided that all the analysis is done at the same load applied to the RVE. 
The same approach can be generalized to cases with more complicated RVEs such as elastoplastic constitutive behavior.

This new computational approach enjoys the following features:
\begin{enumerate}
    \item Only three hyperparameters need to be determined to learn the manifold. And once the data manifold is constructed, minimum computation is required to reconstruct the phase field output. 
    \item There exists an indicator which can pre-estimate the reconstruction error and pre-determine whether an input data is suitable to perform the reconstruction. 
    We would like to emphasize that this feature is very desirable, since compared with more popular machine-learning techniques such as neural networks -- in many of those techniques, it is difficult to predict whether an interpolation is accurate or not without knowing the exact solution.
    \item A number of generalizations can be made, e.g., to three dimensions, and to the types of RVEs, boundary conditions, and outputs. In fact, the output can be of a high dimension, as long as there exists a continuous dependence of the output on the input, which is anyway a prerequisite of a well-posed problem.
    \item The applicability of this approach is promising. The adaptive algorithm makes efficient multiscale fracture simulation possible.
\end{enumerate}


%
\section*{Conflict of interest}
The authors declare that they have no conflict of interest.

\bibliographystyle{spphys}       
\bibliography{Arxiv_Manifold}   

%
%

\end{document}